\documentclass[english, oneside, fontsize=11pt, final, a4paper]{scrartcl}

\pdfoutput=1

\usepackage[T1]{fontenc}
\usepackage[utf8]{inputenc}
\usepackage{babel}

\usepackage{nccmath} 
\usepackage{mathtools} 
\usepackage{amsfonts}
\usepackage{amssymb}
\usepackage{dsfont}
\usepackage{physics}
\usepackage{slashed}
\numberwithin{equation}{section}

\usepackage[final, babel=true]{microtype}
\usepackage{csquotes} 
\usepackage{graphicx}

\makeatletter
\g@addto@macro\bfseries{\boldmath}
\makeatother

\usepackage[nosort]{cite}

\usepackage[table]{xcolor}
\usepackage{hyperref}
\usepackage[ocgcolorlinks]{ocgx2} 
	\hypersetup{
		final,
		unicode=true,
		allcolors={linkcolor},
		pdftitle={The twisted story of worldsheet scattering in η-deformed AdS₅ ⨉ S⁵},
		pdfauthor={Fiona K. Seibold, Stijn J. van Tongeren and Yannik Zimmermann},
		pdfdisplaydoctitle=true,
	}
\usepackage{bookmark}
\usepackage{cleveref}

\usepackage{arydshln}

\newcommand{\resizeToFitPageText}[1]{\resizebox{\textwidth}{!}{\parbox{\textwidth}{#1}}}
\newcommand{\resizeToFitPageMath}[1]{\resizebox{\textwidth}{!}{$#1$}}

\newcommand{\numberthis}{\stepcounter{equation}\tag{\theequation}}

\newcommand{\mup}[1]{{\textnormal{#1}}}
\newcommand{\mdot}{\:.}
\newcommand{\mcomma}{\:,}
\newcommand{\I}{i} 
\newcommand{\E}{e} 

\newcommand{\diag}{\operatorname{diag}}

\DeclareMathOperator{\arsinh}{arsinh}
\DeclareMathOperator{\sgn}{sign}

\newcommand{\ads}{\textup{AdS}_\textup{5}\times\textup{S}^\textup{5}}

\newcommand{\dsS}{\mathbb{S}}
\newcommand{\dsT}{\mathbb{T}}
\newcommand{\scT}{\mathcal{T}}
\newcommand{\identity}{\mathds{1}}

\newcommand\Q[1]{\left[ #1 \right]_q}
\newcommand{\com}[2]{[#1,#2]}

\newcommand{\antipode}{\textsf{S}}
\newcommand{\Alg}{\mathcal}

\newcommand{\SA}{A}
\newcommand{\SB}{B}
\newcommand{\SC}{C}
\newcommand{\SD}{D}
\newcommand{\SE}{E}
\newcommand{\SF}{F}
\newcommand{\SG}{G}
\newcommand{\SH}{H}
\newcommand{\SK}{K}
\newcommand{\SL}{L}

\newcommand{\alternativeCoefficientsStyle}[1]{\hat{#1}}
\newcommand{\aalt}{\alternativeCoefficientsStyle{a}}
\newcommand{\balt}{\alternativeCoefficientsStyle{b}}
\newcommand{\calt}{\alternativeCoefficientsStyle{c}}
\newcommand{\dalt}{\alternativeCoefficientsStyle{d}}

\newcommand{\Ftwist}{\textsf{F}}
\newcommand{\chargeconj}{\mathcal C}

\newcommand{\stringtension}{h} 
\newcommand{\permute}[1]{{#1^\mup{p}}}

\definecolor{linkcolor}{HTML}{00445C} 

\definecolor{color1}{HTML}{A9E6D5} 
\definecolor{color2}{HTML}{F7E8C3} 
\definecolor{color3}{HTML}{B3E4FF} 
\definecolor{color4}{HTML}{FFE7D3} 

\addtokomafont{titlehead}{\raggedleft\ttfamily}
\titlehead{HU-EP-20/17}

\title{The twisted story of worldsheet scattering in $\eta$-deformed $\ads$}
\date{}

\newcommand*{\affaddr}[1]{\normalsize\textit{#1}}
\newcommand*{\affmark}[1]{\textsuperscript{#1}}
\newcommand*{\affmarkpunc}[1]{\hspace{-0.1em}\affmark{#1}}
\newcommand*{\email}[1]{\normalsize\texttt{#1}}

\author{
	Fiona K. Seibold,\affmarkpunc{a}
	Stijn J. van Tongeren,\affmarkpunc{b}
	and
	Yannik Zimmermann\affmark{b}
	\vspace{1cm} \\
	\affmark{a}\affaddr{Institut für Theoretische Physik, ETH Zürich,}\\
	\affaddr{Wolfgang-Pauli-Strasse 27, 8093 Zürich, Switzerland}
	\vspace{0.5cm} \\
	\affmark{b}\affaddr{Institut für Physik, Humboldt-Universität zu Berlin,}\\
	\affaddr{IRIS Gebäude, Zum Grossen Windkanal 6, 12489 Berlin, Germany}
	\vspace{1cm} \\
	\email{fseibold@itp.phys.ethz.ch} \\
	\email{svantongeren@physik.hu-berlin.de} \\
	\email{yannik.zimmermann@physik.hu-berlin.de}
}

\begin{document}

\maketitle

\begin{abstract}
\noindent
We study the worldsheet scattering theory of the $\eta$ deformation of the $\ads$ superstring corresponding to the purely fermionic Dynkin diagram. This theory is a Weyl-invariant integrable deformation of the $\ads$ superstring, with trigonometric quantum-deformed symmetry. We compute the two-body worldsheet S matrix of this string in the light-cone gauge at tree level to quadratic order in fermions. The result factorizes into two elementary blocks, and solves the classical Yang-Baxter equation. We also determine the corresponding exact factorized S matrix, and show that its perturbative expansion matches our tree-level results, once we correctly identify the deformed light-cone symmetry algebra of the string. Finally, we briefly revisit the computation of the corresponding S matrix for the $\eta$ deformation based on the distinguished Dynkin diagram, finding a tree-level S matrix that factorizes and solves the classical Yang-Baxter equation, in contrast to previous results.
\end{abstract}

\pagebreak
\pdfbookmark[section]{\contentsname}{toc} 
\tableofcontents

\section{Introduction}\label{sec:intro}

The discovery and development of integrable structures in the AdS/CFT correspondence has led to impressive insights into quantum field and string theory \cite{Beisert:2010jr,Bombardelli:2016rwb}. On the string theory side the canonical model is the superstring on $\ads$, a maximally supersymmetric sigma model. In recent years integrable deformations of this theory have attracted attention, building on the development of Yang-Baxter sigma models \cite{Klimcik:2002zj,Klimcik:2008eq,Delduc:2013qra}. There is a plethora of Yang-Baxter deformations of the $\ads$ string, with distinct algebraic characteristics and interpretations in terms of string theory and AdS/CFT. We will consider so-called inhomogeneous Yang-Baxter deformations, which algebraically correspond to trigonometric quantum ($q$) deformations \cite{Delduc:2014kha}.\footnote{As the name suggests there are also homogeneous Yang-Baxter deformations \cite{Kawaguchi:2014qwa}, a class which includes e.g.\ the well-known real $\beta$ deformation of the $\ads$ string \cite{Matsumoto:2014nra}. Algebraically these correspond to twisted symmetry \cite{vanTongeren:2015uha,vanTongeren:2018vpb}, see also \cite{Kawaguchi:2013lba}. This twisted symmetry can be used to conjecture field theory duals \cite{vanTongeren:2015uha}.} These deformations are governed by an $R$ operator solving the modified classical Yang-Baxter equation (mCYBE). In the context of the $\ads$ string they are also called $\eta$ deformations.

Studies of the original $\eta$ deformation of the $\ads$ string led to a number of open questions and interesting discoveries. Namely, while Yang-Baxter deformed superstrings have $\kappa$ symmetry \cite{Delduc:2013qra}, the background of the original $\eta$-deformed $\ads$ superstring does not satisfy the supergravity equations of motion \cite{Arutyunov:2015qva}. Rather it satisfies a generalized set of equations \cite{Arutyunov:2015mqj}, which actually derive from $\kappa$ symmetry \cite{Wulff:2016tju}. These equations are believed to guarantee scale invariance, but not Weyl invariance \cite{Arutyunov:2015mqj,Wulff:2016tju,Wulff:2018aku}.\footnote{There have been proposals suggesting that a notion of Weyl invariance may hold for these generalized backgrounds as well \cite{Fernandez-Melgarejo:2018wpg,Muck:2019pwj}. These proposals, however, have troublesome features as discussed in \cite{Muck:2019pwj}.} In order for a Yang-Baxter model background to solve the more restrictive supergravity equations of motion, the $R$ operator generically needs to be unimodular \cite{Borsato:2016ose}.\footnote{Unimodularity is sufficient, while there are subtle counterexamples to necessity, see \cite{Hoare:2016hwh,Wulff:2018aku,Borsato:2018idb}.} This raised the question whether a unimodular inhomogeneous deformation of $\ads$ exists, i.e.\ whether there is a unimodular inhomogeneous solution of the CYBE for $\mathfrak{psu}(2,2|4)$.

The canonical solution of the inhomogeneous CYBE is the so-called Drinfel'd-Jimbo $R$ operator, which is unique for a compact Lie algebra. For noncompact algebras there is freedom corresponding to a choice of simple roots relative to the real form, see \cite{Delduc:2014kha,Hoare:2016ibq} for a discussion in the present context. For superalgebras there is further freedom in whether we choose bosonic or fermionic simple roots, mirroring the lack of uniqueness of Dynkin diagrams for superalgebras. The original $\eta$ deformation \cite{Delduc:2013qra,Arutyunov:2014cda,Arutyunov:2015qva} is based on the Drinfel'd-Jimbo $R$ matrix for the distinguished Dynkin diagram of $\mathfrak{psu}(2,2|4)$, which is not unimodular. Building an $R$ operator relative to the fully fermionic Dynkin diagram instead, gives a unimodular result, and a deformation of $\ads$ that solves the supergravity equations of motion \cite{Hoare:2018ngg}.\footnote{This deformation can also be used as a starting point to generate new homogeneous unimodular deformations by limiting procedures \cite{vanTongeren:2019dlq}.} We will refer to these two distinct deformations as the distinguished and fermionic ($\eta$) deformations respectively. The metric and B field of these models are equal, while their dilatons and Ramond-Ramond (RR) forms differ.\footnote{There are unimodular deformations that one can obtain from the one of \cite{Hoare:2018ngg} by permutations of the bosonic roots as in \cite{Delduc:2014kha,Hoare:2016ibq}. Here we focus the case which gives the \enquote{standard} metric and B field with magnetic $H$ flux.}

In this paper we will be investigating the worldsheet scattering theory for the fermionic deformation. There are concrete open questions motivating our study, in addition to broader interest in the quantum integrable structure of this Weyl invariant, integrable deformation of the $\ads$ string, with trigonometric $q$-deformed symmetry. Namely, the scattering theory of the distinguished $\eta$ deformation shows some interesting features that we would like to contrast with the corresponding fermionic ones. First, the tree-level S matrix for the distinguished model was found not to satisfy the classical Yang-Baxter equation (CYBE) \cite{Arutyunov:2015qva}, while the model is classically integrable. A non-local two-particle change of scattering states was required to restore this hallmark requirement of integrability, as well as to match the expansion of the exact factorized $\mathfrak{su}_q(2|2)_\mup{c.e.}^{\oplus2}$ S matrix \cite{Beisert:2008tw,Hoare:2011wr} expected to describe this model. This unexpected friction between classical integrability and tree-level factorized scattering, and the subtle redefinition of scattering states, could be related to the lack of Weyl invariance of this model, which is restored for the fermionic deformation.\footnote{In general we would expect Weyl invariance to come into play only at loop level, however.} Second, the distinguished deformed model displays so-called \enquote{mirror duality} \cite{Arutynov:2014ota,Arutyunov:2014cra,Arutyunov:2014jfa,Pachol:2015mfa} at the bosonic level and in terms of its conjectured exact $S$ matrix. In short, in the light-cone gauge fixed theory, inversion of the deformation parameter is equivalent to a double Wick rotation on the worldsheet, which curiously relates the thermodynamic and spectral properties of the model. Studying the S matrix for the fermionic deformation is a first step towards investigating similar properties here.


We study two aspects of the worldsheet scattering theory of the fermionic $\eta$ deformation of the $\ads$ string. First, we compute the two body S matrix perturbatively at tree level with up to two fermions. We find that the resulting $\dsT$ matrix solves the CYBE, in line with expected integrability. The $\dsT$ matrix factorizes, and we expect the factors to be related to an exact S matrix for $\mathfrak{su}_q(2|2)_\mup{c.e.}$, analogously to the undeformed and distinguished deformed string. However, only the distinguished $\mathfrak{su}_q(2|2)_\mup{c.e.}$ S matrix is explicitly known \cite{Beisert:2008tw}. As such, second we determine the form of the exact $\mathfrak{su}_q(2|2)_\mup{c.e.}$ S matrix for the fermionic deformation. We do this by taking advantage of a twist relating the Hopf algebras underlying the distinguished and fermionic deformations of $\mathfrak{sl}_q(2|2)_\mup{c.e.}$. Next, based on the embedding of the two copies of $\mathfrak{su}(2|2)$ in $\mathfrak{psu}(2,2|4)$, we conjecture that the deformation of the off-shell light-cone symmetry algebra of the string takes the form $\mathfrak{su}_{1/q}(2|2)_\mup{c.e.}\oplus\mathfrak{su}_{q}(2|2)_\mup{c.e.}$. Semi-classically $q=e^{-\kappa/\stringtension}$, where $\kappa$ is the deformation parameter in the action, and $\stringtension$ is the string tension. The associated exact S matrix is of the form  $S_0\, S(1/q)\otimes S(q)$, where $S_0$ is a scalar prefactor. The perturbative expansion of this exact S matrix matches our tree-level $\dsT$ matrix.

We originally benchmarked our computations of the perturbative S matrix on the undeformed $\ads$ string. After we obtained our results for the fermionic deformation we decided to also run through the distinguished background given in \cite{Arutyunov:2015qva}. Unexpectedly, in contrast to \cite{Arutyunov:2015qva} we find a perturbative S matrix that directly solves the CYBE, and factorizes in line with the distinguished $\mathfrak{su}_q(2|2)_\mup{c.e.}$ S matrix. In this case the S matrix factors are such that an inversion of the deformation parameter is equivalent to a change of basis, and there is effectively no distinction between $S_0\, S(1/q)\otimes S(q)$ and $S_0\, S(q)^{\otimes2}$.

This paper is organized as follows. In the next section we discuss the string Lagrangian, its gauge fixing, and its expansion in powers of fields. Then in section \ref{sec:treelevelcomputation} we compute the associated tree-level S matrix, and discuss its factorized structure. In section \ref{sec:exactSmatrix} we review the construction of the distinguished $\mathfrak{su}(2|2)_\mup{c.e.} $ S matrix, and twist this construction to find the fermionic exact S matrix. We then analyze the structure of the light-cone symmetry algebra in section \ref{sec:comparison}, and show that the expansion of the corresponding exact S matrix matches our tree-level computation. In section \ref{sec:distinguishedcase} we discuss our results regarding the distinguished case. Finally we conclude and list several open questions. We provide appendices on our spinor conventions, our implementation of the Feynman diagram computations, and a translation of $\mathfrak{su}(2|2)$ $R$ operators in the sigma model and exact $S$ matrix computations.

\section{Deformed Lagrangian}
\label{sec:Lagrangian}

To compute the tree-level two-body worldsheet S matrix of the fermionic $\eta$ deformed string in the light-cone gauge, we need the corresponding action in the light-cone gauge, expanded to quartic order in the fields. Rather than working directly with the Yang-Baxter sigma model action \cite{Delduc:2013qra}, we will work with the standard Green-Schwarz (GS) action and substitute the background for the fermionic deformation found in \cite{Hoare:2018ngg}.


\subsection{The GS string to second order in fermions}

Written out, the Lagrangian for a type IIB GS superstring in a generic background, to second order in the fermions, takes the form\footnote{See e.g.\ \cite{Wulff:2013kga}, but note that we use a different sign convention on the fermionic worldsheet $\epsilon$ term, in line with \cite{Arutyunov:2015qva}.}
\begin{equation}
\label{eq:Lagrangian}
\begin{aligned}
\mathcal{L} = {} & \sqrt{-\mbox{det}_{\gamma\delta}\, h_{\gamma \delta}}\,h^{\alpha\beta} \hat{g}_{MN} \partial_\alpha x^M \partial_\beta x^N - \epsilon^{\alpha\beta} \hat{B}_{MN} \partial_\alpha x^M \partial_\beta x^N\\
& + i\sqrt{-\mbox{det}_{\gamma\delta}\, h_{\gamma \delta}}\,h^{\alpha\beta} \partial_\alpha x^M \bar{\theta} \Gamma_M \partial_\beta\theta + i\epsilon^{\alpha \beta} \partial_\alpha x^M \bar{\theta} \Gamma_M \sigma_3 \partial_\beta \theta \,,
\end{aligned}
\end{equation}
where $\theta = (\theta_1,\theta_2)$ is a doublet of 10D Majorana-Weyl spinors, with the Pauli matrix $\sigma_3$ acting in this two dimensional space. The worldsheet metric $h_{\alpha \beta}$ has signature $(-1,1)$, and $\epsilon^{\tau\sigma} =1$. In this expression we have combined certain fermionic terms with the bosonic metric $g$ and B field $B$, i.e.
\begin{equation}
\label{eq:Bhatandghatdef}
\begin{aligned}
\hat{g}_{MN} &\, = g_{MN} - \tfrac{i}{4} \bar{\theta} \Gamma_{(M} \slashed{\omega}_{N)} \theta + \tfrac{i}{8} \bar{\theta} \Gamma_{(M}H_{N)PQ}\Gamma^{PQ} \sigma_3 \theta + \tfrac{i}{8} \bar{\theta} \Gamma_{(M} \mathcal{S} \Gamma_{N)} \theta,
 \\
\hat{B}_{MN} &\, = B_{MN} + \tfrac{i}{4} \bar{\theta} \Gamma_{[M} \slashed{\omega}_{N]}\sigma_3 \theta  - \tfrac{i}{8} \bar{\theta} \Gamma_{[M}H_{N]PQ} \Gamma^{PQ}\theta - \tfrac{i}{8} \bar{\theta} \sigma_3 \Gamma_{[M} \mathcal{S} \Gamma_{N]} \theta,
\end{aligned}
\end{equation}
with round and rectangular brackets denoting symmetrization and antisymmetrization respectively, defined with the usual factor of $1/n!$. Here $\omega$ denotes the spin connection, $H=dB$, and\footnote{Slashes denote contraction with the appropriate set of $\Gamma$ matrices: $\slashed{A} \equiv A_{M\ldots L} \Gamma^M\ldots \Gamma^L$.}
\begin{equation}
\label{eq:Sformdefinition}
\mathcal{S} = - \left(\epsilon \slashed{\mathcal{F}}^{(1)} + \tfrac{1}{3!} \sigma_1 \slashed{\mathcal{F}}^{(3)} + \tfrac{1}{2\cdot 5!}\epsilon \slashed{\mathcal{F}}^{(5)}\right),
\end{equation}
where $\epsilon \equiv i\sigma_2$. Assuming a dilaton exists, $\slashed{\mathcal{F}}$ encodes the RR forms and dilaton via $\mathcal{F}^{(n)} =e^\Phi F^{(n)}$.

\subsection{Light-cone gauge fixing}

We assume that our general background has two isometries $t$ and $\phi$, where $t$ is timelike and $\phi$ is spacelike, and introduce the light-cone coordinates
\begin{equation}
x^+ = \frac{1}{2}\left(t + \phi\right), \quad x^- = \phi - t.
\end{equation}
The uniform light-cone gauge then consists of fixing
\begin{equation}
x^+=\tau, \quad p_-=1,
\end{equation}
where $\tau$ is the worldsheet time and $p_-$ the momentum conjugate to $x^-$. We can shortcut gauge fixing in the Hamiltonian framework by noting that momentum and winding interchange under T duality, so that if we formally T dualize our model in $x^-$, calling the dual direction $\psi$, our uniform light-cone gauge condition becomes
\begin{equation}
x^+ = \tau, \quad \psi = \sigma.
\end{equation}
Upon integrating out the worldsheet metric the T dualized Lagrangian takes the square root form typical for a light-cone gauge. In this picture the gauge condition can be directly substituted in the Lagrangian. This light-cone gauge fixing should be accompanied by a corresponding $\kappa$-symmetry gauge choice for the fermions of the form
\begin{equation}
\Gamma^p \theta = 0,
\end{equation}
where $\Gamma^p$ is the tangent space counterpart of $\Gamma^+$, defined in \eqref{eq:tangentspacelightconeGammas} for our particular case. We assume this gauge fixing from here on, see e.g.\ \cite{Arutyunov:2014jfa} for further details.\footnote{The possibility of gauge fixing via T duality was originally observed for the $\ads$ string in \cite{Kruczenski:2004cn}.}

To simplify expressions we introduce the T-dual metric $\mathring{g}$, B field $\mathring{B}$, and gamma matrices $\mathring{\Gamma}$
\begin{equation}
\begin{alignedat}{2}
	\mathring{g}_{\psi\psi} = \frac{1}{\hat{g}_{--}}, \quad
	\mathring{g}_{\psi \bar{M} } & =-\frac{\hat{B}_{-\bar{M}}}{\hat{g}_{--}}, \quad &
	\mathring{g}_{\bar{M}\bar{N}} & =\hat{g}_{\bar{M}\bar{N}} -  \frac{\hat{g}_{-\bar{M}}\hat{g}_{-\bar{N}}-\hat{B}_{-\bar{M}}\hat{B}_{-\bar{N}}}{\hat{g}_{--}},
	\\
	\mathring{B}_{\psi \bar{M} } & =-\frac{\hat{g}_{-\bar{M}}}{g_{--}}, \quad &
	\mathring{B}_{\bar{M}\bar{N}} & =\hat{B}_{\bar{M}\bar{N}} -  \frac{\hat{g}_{-\bar{M}}\hat{B}_{-\bar{N}}-\hat{B}_{-\bar{M}}\hat{g}_{-\bar{N}}}{\hat{g}_{--}},
	\\
	\mathring{\Gamma}_\psi & = \frac{1}{\hat{g}_{--}} \Gamma_-,\quad &
	\mathring{\Gamma}_{\bar{M}} & = \Gamma_{\bar{M}} - \frac{g_{-\bar{M}}}{g_{--}} \Gamma_{-},
\end{alignedat}
\end{equation}
where $\bar{M}$ and $\bar{N}$ run over the coordinates not involved in the T duality. For $\mathring{g}$ and $\mathring{B}$ the right hand side of these equations is implicitly expanded to second order in fermions. With these definition the general gauge fixed action to quadratic order in fermions takes the form
\begin{equation}
\mathcal{L}^\mup{g.f.} = 2 \sqrt{-G} + E,
\end{equation}
where $G = \det_{\alpha\beta} G_{\alpha\beta}$ and $E = \epsilon^{\alpha \beta} E_{\alpha\beta}$ with
\begin{equation}
\begin{aligned}
G_{\alpha\beta} &= \mathring{g}_{MN}\partial_\alpha x^M \partial_\beta x^N + i \partial_{(\alpha|} x^{\bar{M}} \bar{\theta} \mathring{\Gamma}_{\bar{M}} \partial_{|\beta)} \theta + i \partial_{(\alpha|} \psi \bar{\theta} \mathring{\Gamma}_\psi \sigma_3 \partial_{|\beta)} \theta ,\\
E_{\alpha\beta} &= - \mathring{B}_{MN} \partial_\alpha x^M \partial_\beta x^N + i \partial_\alpha \psi \bar{\theta} \mathring{\Gamma}_\psi \partial_\beta \theta + i\partial_\alpha x^{\bar{M}}\bar{\theta} \mathring{\Gamma}_{\bar{M}} \sigma_3 \partial_\beta \theta,
\end{aligned}
\end{equation}
again implicitly expanded to second order in fermions, and evaluated on the gauge fixing condition $x^+ = \tau, \psi = \sigma$. Here the indices $M$ and $N$ run over $x^+,\psi$ and the transverse fields, and round brackets on indices indicate symmetrization. The gauge-fixed string action is
\begin{equation}
\label{eq:gfaction}
S = -\frac{\stringtension}{2} \int \dd[2]{\sigma}  \mathcal{L}^\mup{g.f.} =  - \stringtension \int \dd[2]{\sigma} \sqrt{-G} + \frac{1}{2} E,
\end{equation}
where $\stringtension$ is the string tension. When we take the string tension into account in the T duality and gauge fixing, consistency of $\psi = \sigma$ with $p_-=1$, fixes the string length to be $P_-/\stringtension$, where $P_-$ is the integrated charge associated to $p_-$, see e.g.\ \cite{Zarembo:2009au,Arutyunov:2014jfa} for details.\footnote{Before substituting the gauge condition, our Nambu-Goto type action is manifestly reparametrization invariant, so we can freely rescale $\sigma$. This rescaling remains a symmetry upon gauge fixing if we correspondingly adapt the gauge condition on $\psi$.}

\subsection{\texorpdfstring{$\eta$-deformed $\ads$}{η-deformed AdS₅ ⨉ S⁵}}

The metric and B field for our fermionic deformation are the same as the ones for the distinguished deformation, given by \cite{Arutyunov:2013ega}\footnote{The authors of \cite{Arutyunov:2013ega} use trigonometric coordinates $\zeta$ and $\xi$ related to our $x$ and $w$ as $x = \sin \zeta$, $w=\sin \xi$.}
\begin{equation*}
\resizeToFitPageMath{
\begin{aligned}
\dd s^2 &=
\frac{1}{1-\kappa^2\rho^2}\left(-(1+\rho^2) \dd t^2 + \frac{\dd \rho^2}{1+\rho^2}\right) + \frac{\rho^2}{1+\kappa^2\rho^4 x^2} \left( (1-x^2) \dd \psi_1^2+ \frac{\dd x^2 }{1-x^2}\right) + \rho^2 x^2 \dd \psi_2^2 \\
& \quad + \frac{1}{1+\kappa^2r^2}\left((1-r^2) \dd \phi^2 + \frac{\dd r^2}{1-r^2}\right) + \frac{r^2}{1+\kappa^2r^4 w^2} \left( (1-w^2) \dd \phi_1^2+ \frac{\dd w^2 }{1-w^2} \right) + r^2 w^2 \dd \phi_2^2 \mcomma \\
B & = \frac{\kappa \rho}{1-\kappa^2 \rho^2} \dd t \wedge \dd \rho + \frac{\kappa \rho^4 x}{1+\kappa^2 \rho^4 x^2} \dd \psi_1 \wedge \dd x + \frac{\kappa r}{1+\kappa^2 r^2} \dd \phi \wedge \dd r -\frac{\kappa r^4 w}{1+\kappa^2 r^4 w^2} \dd \phi_1 \wedge \dd w \mcomma
\end{aligned}
}
\end{equation*}
where $\kappa$ is the deformation parameter. The RR sector of the fermionic deformed model has a nonzero three form and a nonzero five form. As the expressions are large, we refer to the original paper \cite{Hoare:2018ngg} instead of reproducing the RR forms here.

Our conventions for light-cone gauge fixing and the computation of the perturbative S matrix for this background are analogous to those for the undeformed model, see e.g.\ the review \cite{Arutyunov:2009ga}. The two coordinates labeled $t$ and $\phi$ in the background above are isometric, and are the coordinates used in the light-cone gauge fixing. To get the interaction Lagrangian for the perturbative S matrix we first change to a different basis of transverse fields denoted $z_i$, $i=1,\ldots,4$ and $y_j$, $j=1,\ldots,4$. These are related to the transverse coordinates used above as
\begin{equation}
\begin{aligned}
\frac{z_1 + i z_2}{1-\tfrac{1}{4}z^2} & = \rho \sqrt{1-x^2} e^{i\psi_1}\, ,  \quad\frac{z_3 + i z_4}{1-\tfrac{1}{4}z^2}=\rho \,x\, e^{i\psi_2} \, , \quad z^2\equiv z_i^2\, ,\\
\frac{y_1 + i y_2}{1+\tfrac{1}{4}y^2} & = r\sqrt{1-w^2} e^{i\phi_1}\, , \quad\frac{y_3 + i y_4}{1+\tfrac{1}{4}y^2}=r \,w\, e^{i\phi_2} \, ,\quad y^2\equiv y_i^2 \, .
\end{aligned}
\end{equation}
In what follows we have (implicitly) applied this coordinate redefinition to the background, including the RR fields. We fix our spinor conventions in terms of these new coordinates directly, as discussed in appendix \ref{app:spinorconventions}.

\subsection{Expansion of the action}
\label{sec:expansion-of-action}

For the computation of the tree-level two-body S matrix we need the gauge-fixed action to quartic order in the transverse fields, keeping in mind that we restricted to quadratic order in fermions from the start. Physically we
consider the string action \eqref{eq:gfaction}, rescale the transverse fields by $1/\sqrt{\stringtension}$, e.g.\ $z_1 \rightarrow z_1/\sqrt{\stringtension}$, and keep terms up to order $1/\stringtension$, i.e.\footnote{In the approach of the review \cite{Arutyunov:2009ga} $\sigma$ is rescaled by $\stringtension$ to remove explicit dependence on $\stringtension$ from the gauge fixed Hamiltonian. Our conventions and starting point circumvent this, but of course in both cases we end up with a string length of $P_-/\stringtension$ and only an overall factor of $\stringtension$ before expanding.}
\begin{equation}
	S = \int \dd[2]{\sigma} \left( \mathcal{L}_2 + \frac{1}{\stringtension} \mathcal{L}_4 + \ldots\right),
\end{equation}
where by convention we have absorbed a sign in the definition of $\mathcal{L}_{2,4}$. This expansion is straightforward but computationally involved due to the complicated nature of the backgrounds.\footnote{To give some technical details, we evaluated the gauge-fixed Lagrangian described above, formally expanding in fermions whenever possible before substituting concrete expressions. We expressed everything in terms of the bosonic coordinates, the two gauge fixed spinors $\theta_{1}$ and $\theta_2$, and a set of canonically ordered abstract tangent space gamma matrices. We discarded any terms that are zero due to the $\kappa$-gauge fixing, expanded the resulting expressions to appropriate order in bosons, and finally substituted concrete spinors and gamma matrices. In practice we were not able to sufficiently simplify the coordinate transformed RR forms before expanding, so we resorted to expanding the contributions of the RR forms to second order in the bosons before substituting them in the gauge-fixed Lagrangian. }

At the quadratic level we find
\begin{equation}
	\begin{aligned}
		\mathcal{L}_2 = {} &
			\epsilon_{a b} \epsilon_{\dot{a} \dot{b}} \qty(
				- \partial_\tau Y^{a \dot{a}} \partial_\tau Y^{b \dot{b}}
				+ \partial_\sigma Y^{a \dot{a}} \partial_\sigma Y^{b \dot{b}}
				+ (1 + \kappa^2) Y^{a \dot{a}} Y^{b \dot{b}}
			)
			\\ &
			+ \epsilon_{\alpha \beta} \epsilon_{\dot{\alpha} \dot{\beta}} \qty(
				- \partial_\tau Z^{\alpha \dot{\alpha}} \partial_\tau Z^{\beta \dot{\beta}}
				+ \partial_\sigma Z^{\alpha \dot{\alpha}} \partial_\sigma Z^{\beta \dot{\beta}}
				+ (1 + \kappa^2) Z^{\alpha \dot{\alpha}} Z^{\beta \dot{\beta}}
			)
			\\ &
			+ \I \theta^\dagger_{a \dot{\alpha}} \partial_\tau \theta^{a \dot{\alpha}}
			+ \frac{1}{2} \qty(
				\epsilon_{a b} \epsilon_{\dot{\alpha} \dot{\beta}} \theta^{a \dot{\alpha}} \partial_\sigma \theta^{b \dot{\beta}}
				- \epsilon^{a b} \epsilon^{\dot{\alpha} \dot{\beta}} \theta^\dagger_{a \dot{\alpha}} \partial_\sigma \theta^\dagger_{b \dot{\beta}}
			)
			- \sqrt{1 + \kappa^2} \theta^\dagger_{a \dot{\alpha}} \theta^{a \dot{\alpha}}
			\\ &
			+ \I \eta^\dagger_{\alpha \dot{a}} \partial_\tau \eta^{\alpha \dot{a}}
			+ \frac{1}{2} \qty(
				\epsilon_{\alpha \beta} \epsilon_{\dot{a} \dot{b}} \eta^{\alpha \dot{a}} \partial_\sigma \eta^{\beta \dot{b}}
				- \epsilon^{\alpha \beta} \epsilon^{\dot{a} \dot{b}} \eta^\dagger_{\alpha \dot{a}} \partial_\sigma \eta^\dagger_{\beta \dot{b}}
			)
			- \sqrt{1 + \kappa^2} \eta^\dagger_{\alpha \dot{a}} \eta^{\alpha \dot{a}}\mcomma
	\end{aligned}
\end{equation}
where we have introduced the complex fields $Y$ and $Z$ via
\begin{equation}
	\label{eq:zytoZY}
	\begin{aligned}
		\qty(Y^{a \dot{a}})
		& =
		\begin{pmatrix}
			Y^{2\dot{2}} &
			-Y^{2\dot{1}} \\
			Y^{1\dot{2}} &
			-Y^{1\dot{1}}
		\end{pmatrix}
		= \frac{1}{2}
		\begin{pmatrix}
			y_3 - i y_4 &
			-y_1 +i y_2 \\
			y_1 + i y_2 &
			y_3 + i y_4
		\end{pmatrix}
		\mcomma \\
		\qty(Z^{\alpha \dot{\alpha}})
		& =
		\begin{pmatrix}
			Z^{3\dot{3}} &
			-Z^{3\dot{4}} \\
			Z^{4\dot{3}} &
			-Z^{4\dot{4}}
		\end{pmatrix}
		= \frac{1}{2}
		\begin{pmatrix}
			z_3 - i z_4 &
			-z_1 +i z_2 \\
			z_1 + i z_2 &
			z_3 + i z_4
		\end{pmatrix}
		\mcomma
	\end{aligned}
\end{equation}
in addition to the fermions $\theta^{a \dot{\alpha}}$ and $\eta^{\alpha \dot{a}}$ parametrizing the spinors, as presented in \cref{spinor-parametrization} in the appendix.
The indices on these fields label their transformations with respect to the $\mathfrak{su}(2)^{\oplus 4}$ symmetry of the undeformed model, acting from the left and the right on the matrices.
We denote the indices 1 and 2 with Latin letters ($a$ and $b$) and the indices 3 and 4 with Greek letters ($\alpha$ and $\beta$).
For an index running from 1 to 4 we use capital Latin letters $\scriptstyle{M},\scriptstyle{N},\dots$.
The Levi-Civita symbols $\epsilon_{a b}$ and $\epsilon_{\alpha \beta}$ are defined for Latin and Greek indices individually, i.e.\ $\epsilon_{12} = \epsilon^{12} = 1$ and $\epsilon_{34} = \epsilon^{34} = 1$.

The reality of $y_i$ and $z_i$ implies the reality condition
\begin{equation}
	\label{reality-bosons}
	\qty(Y^{a \dot{a}})^\dagger = - \epsilon_{a b} \epsilon_{\dot{a} \dot{b}} Y^{b \dot{b}}
	\mcomma \qquad
	\qty(Z^{\alpha \dot{\alpha}})^\dagger = - \epsilon_{\alpha \beta} \epsilon_{\dot{\alpha} \dot{\beta}} Z^{\beta \dot{\beta}}
	\mdot
\end{equation}
so that from the worldsheet perspective the model contains 8 real scalar bosons and 8 complex scalar fermions (Grassmann fields), all with mass $\sqrt{1 + \kappa^2}$.
The interaction Lagrangian $\mathcal{L}_4$ is too large to be meaningfully presented here, but can be found in the Mathematica notebook attached to the arXiv submission of this paper.

We note that our conventions at this point differ from those of~\cite{Arutyunov:2013ega,Arutyunov:2015qva} and the review~\cite{Arutyunov:2009ga}.
Firstly, with respect to~\cite{Arutyunov:2013ega,Arutyunov:2015qva} we interchanged indices $\dot{1} \leftrightarrow \dot{2}$ and $\dot{3} \leftrightarrow \dot{4}$ for convenient comparison to the exact S matrix later.
Similarly, with respect to the review~\cite{Arutyunov:2009ga} we interchanged $1 \leftrightarrow 2$ and $\dot{3} \leftrightarrow \dot{4}$,
which is a symmetry of the undeformed model.
Secondly, the authors of \cite{Arutyunov:2013ega,Arutyunov:2015qva} parametrized the string tension as $\stringtension = g \sqrt{1+\kappa^2}$, and rescaled the fields by $1/\sqrt{g}$ rather than $1/\sqrt{\stringtension}$. Hence our and their interaction terms, had \cite{Arutyunov:2013ega,Arutyunov:2015qva} worked in a Lagrangian framework, are related as
\begin{equation}
\mathcal{L}_{4}(\varphi) = (1+\kappa^2) \bar{\mathcal{L}}_{4}(\bar{\varphi}).
\end{equation}
where we denote quantities from \cite{Arutyunov:2013ega,Arutyunov:2015qva} with bars, with $\varphi$ collectively denoting the rescaled transverse fields. Moreover, in light-cone gauge fixing we implicitly rescale $\sigma$ by $1/\stringtension$ compared to the implicit rescaling by $1/g$ of \cite{Arutyunov:2013ega,Arutyunov:2015qva}. As a result
\begin{equation}
\sigma= \frac{1}{\sqrt{1+\kappa^2}}\, \bar{\sigma} \quad \implies \quad p = \sqrt{1+\kappa^2}\, \bar{p},
\end{equation}
where $p$ is the spatial worldsheet momentum used in the S matrix below. Under these identifications, our quadratic Lagrangian matches the one of \cite{Arutyunov:2015qva}. Our bosonic interaction Lagrangian should correspond to the bosonic interactions of \cite{Arutyunov:2013ega,Arutyunov:2015qva} in the Hamiltonian setting, while the fermionic interaction terms based on different RR sectors are inherently different.

\section{Perturbative S matrix}
\label{sec:treelevelcomputation}

With our kinetic and interaction Lagrangians we are ready to compute the tree-level S matrix using Feynman diagram methods.
We first present our choice of mode expansion used to determine the asymptotic scattering states, then give the result for the T matrix, and finally show that this result factorizes, albeit in a form that slightly deviates from the expectations from the distinguished case of \cite{Arutyunov:2015qva}. We give a detailed discussion of this factorized structure and its relation to the exact result in \cref{sec:comparison}.

\subsection{On-shell mode expansion}
\label{sec-mode-expansion}

For the in- and out-states of the Feynman amplitudes we need the solutions of the equations of motion for $\mathcal{L}_2$. These are given by the on-shell mode expansions%
\footnote{
	Note that in the limit $\kappa \to 0$ the review \cite{Arutyunov:2009ga} gives an expansion that differs by factors of $\pm\I$ for the fermions.
	This would give a T matrix that differs by some (physically inconsequential) signs from the T matrix of \cite{Klose:2006zd}, which is the one reproduced in \cite{Arutyunov:2009ga}.
}
\begin{fleqn}
\begin{equation}
	\label{mode-expansion}
	\mathrlap{
	\begin{aligned}
		Y^{a \dot{a}}(\tau, \sigma) =
		\frac{1}{\sqrt{2\pi}} \int \dd{p} \frac{1}{2 \sqrt{\omega_p}}
		& \qty(
			+ \E^{\I (p \sigma - \omega_p \tau)} a^{a \dot{a}}(p)
			- \E^{-\I (p \sigma - \omega_p \tau)} \epsilon^{a b} \epsilon^{\dot{a} \dot{b}} a^\dagger_{b \dot{b}}(p)
		)
		\mcomma
		\\
		Z^{\alpha \dot{\alpha}}(\tau, \sigma) =
		\frac{1}{\sqrt{2\pi}} \int \dd{p} \frac{1}{2 \sqrt{\omega_p}}
		& \qty(
			+ \E^{\I (p \sigma - \omega_p \tau)} a^{\alpha \dot{\alpha}}(p)
			- \E^{-\I (p \sigma - \omega_p \tau)} \epsilon^{\alpha \beta} \epsilon^{\dot{\alpha} \dot{\beta}} a^\dagger_{\beta \dot{\beta}}(p)
		)
		\mcomma
		\\
		\theta^{a \dot{\alpha}}(\tau, \sigma) =
		\frac{\E^{-\I \pi/4}}{\sqrt{2\pi}} \int \dd{p} \frac{1}{\sqrt{\omega_p}}
		& \qty(
			- \I \E^{\I (p \sigma - \omega_p \tau)} f_p^* a^{a \dot{\alpha}}(p)
			- \I \E^{-\I (p \sigma - \omega_p \tau)} h_p^* \epsilon^{a b} \epsilon^{\dot{\alpha} \dot{\beta}} a^\dagger_{b \dot{\beta}}(p)
		)
		\mcomma
		\\
		\eta^{\alpha \dot{a}}(\tau, \sigma) =
		\frac{\E^{-\I \pi/4}}{\sqrt{2\pi}} \int \dd{p} \frac{1}{\sqrt{\omega_p}}
		& \qty(
			+ \I \E^{\I (p \sigma - \omega_p \tau)} f_p a^{\alpha \dot{a}}(p)
			+ \I \E^{-\I (p \sigma - \omega_p \tau)} h_p \epsilon^{\alpha \beta} \epsilon^{\dot{a} \dot{b}} a^\dagger_{\beta \dot{b}}(p)
		)
		\mcomma
	\end{aligned}
}
\end{equation}
\end{fleqn}
where in comparison to the distinguished case of \cite{Arutyunov:2015qva} it is more convenient to use $f_p^*$ and $h_p^*$ for $\theta^{a \dot{\alpha}}$, as this enables a direct comparison with the exact result of \cref{sec:exactSmatrix}.
The dispersion relation is
\begin{equation}
\label{eq:dispersion}
	\omega_p = \sqrt{1 + \kappa^2 + p^2}
	\mcomma
\end{equation}
and the wave functions for the fermions are given by
\begin{gather}
	f_p = \frac{\sqrt{p + \I \kappa}}{\sqrt{p - \I \kappa}} \sqrt{\frac{\omega_p + \sqrt{1+\kappa^2}}{2}}
	\mcomma
	\qquad
	h_p = \frac{p}{2 f_p}
	\mcomma
	\\
	\abs{f_p}^2 - \abs{h_p}^2 = \sqrt{1+\kappa^2}
	\mcomma
	\qquad
	\abs{f_p}^2 + \abs{h_p}^2 = \omega_p
	\mdot
\end{gather}
We reformulated $f_p$ from \cite{Arutyunov:2015qva} such that we obtain manifestly continuous amplitudes for all $p_1 > p_2$ when choosing the standard branch for the square root function.%
\footnote{
	\label{explanation-phase-f}%
	The equations of motion fix $f_p$ and $h_p$ only up to a phase factor
	-- in parallel to the phase freedom of the parameter $\gamma$ of the exact S matrix (cf.\ \cref{explanation-phase-gamma}).
	This reflects the fact that the phase of the scattering amplitudes is physically insignificant.
	If we want the amplitudes to be real, there are two natural choices for the phase factor:
	Firstly, the choice of \cite{Arutyunov:2015qva}, which exactly matches the undeformed result but is discontinuous.
	Secondly, our choice, which is continuous but for certain values of $p_1$ and $p_2$ only matches the undeformed result up to a (physically inconsequential) sign.
	We chose the latter, favoring continuous amplitudes, and only noticed the sign mismatch after finishing publishing this paper.
	A third -- and probably most appropriate -- option would be to drop the requirement of realness for the amplitudes and move the phase factor $\frac{\sqrt{p + \I \kappa}}{\sqrt{p - \I \kappa}}$ from $f_p$ to $\gamma$.
	This produces complex and continuous amplitudes that exactly match the undeformed result.
}

Upon quantization we have $(a^{M \dot{N}})^\dagger = a^\dagger_{M \dot{N}}$ for all operators.
For the bosons this stems from the reality condition \eqref{reality-bosons},
for the fermions it is a result of the equations of motion.
It reduces the number of degrees of freedom on-shell effectively from 8 complex to 8 real scalar fermions.

\subsection{T matrix}
\label{T-matrix}
We are going to calculate the $2 \to 2$ scattering matrix $\dsS$ from the gauge fixed, deformed and expanded Lagrangian of \cref{sec:expansion-of-action}.
For this, we expand $\dsS$ in terms of the tree-level $\dsT$ matrix as
\begin{equation}
	\dsS = \identity + \frac{\I}{\stringtension} \dsT + \dots
\end{equation}
and follow the standard Feynman diagram procedure, adapted to some of the intricacies of our model --
details are presented in \cref{feynman-diagrammatics}.
The scattering process depends on two momenta, $p_1$ and $p_2$, with $p_1 > p_2$ by assumption.
The scattering states are
$\ket*{a^\dagger_{M \dot{N}}(p_1) a^\dagger_{P \dot{Q}}(p_2)}
= a^\dagger_{M \dot{N}}(p_1) a^\dagger_{P \dot{Q}}(p_2) \ket{0}$.
We label these states by their particle content and have the first and second particle depend on $p_1$ and $p_2$ respectively.
For example we write
\begin{equation}
	\ket{Y_{a \dot{a}} \theta_{b \dot{\beta}}}
	\equiv
	\ket{a^\dagger_{a \dot{a}}(p_1) a^\dagger_{b \dot{\beta}}(p_2)}
	\mcomma \qquad
	\ket{Z_{\alpha \dot{\alpha}} \eta_{\beta \dot{b}}}
	\equiv
	\ket{a^\dagger_{\alpha \dot{\alpha}}(p_1) a^\dagger_{\beta \dot{b}}(p_2)}
	\mdot
\end{equation}
The T matrix is given in the following by its action on the two-particle states.

\newcommand{\dotted}[1]{\mathring{#1}}
\newcommand{\barred}[1]{\bar{#1}}

\newcommand{\Aalgebraic}{\mathcal{A}}
\newcommand{\Balgebraic}{\mathcal{B}}
\newcommand{\Galgebraic}{\mathcal{G}}
\newcommand{\Walgebraic}{\mathcal{W}}
\newcommand{\Calgebraic}{\mathcal{C}}
\newcommand{\Halgebraic}{\mathcal{H}}
\newcommand{\CbarAlgebraic}{\barred{\Calgebraic}}
\newcommand{\HbarAlgebraic}{\barred{\Halgebraic}}

\newcommand{\Cundot}{C}
\newcommand{\Cdot}{\Calgebraic}
\newcommand{\CbarUndot}{\barred{C}}
\newcommand{\CbarDot}{\CbarAlgebraic}

\newcommand{\Hundot}{H}
\newcommand{\Hdot}{\Halgebraic}
\newcommand{\HbarUndot}{\barred{H}}
\newcommand{\HbarDot}{\HbarAlgebraic}

\begingroup
\newcommand{\myintertext}[1]{\intertext{\vskip 5pt \noindent \bfseries \sffamily #1\vspace{10pt}}}
\begin{align*}
\myintertext{Boson-Boson}
	\dsT \ket*{Y_{a \dot{a}} Y_{b \dot{b}}} = {}&
		+ 2 \Aalgebraic \ket*{Y_{a \dot{a}} Y_{b \dot{b}}}
		+ (\Balgebraic + \Walgebraic \epsilon_{\dot{a} \dot{b}}) \ket*{Y_{a \dot{b}} Y_{b \dot{a}}}
		+ (\Balgebraic - \Walgebraic \epsilon_{a b}) \ket*{Y_{b \dot{a}} Y_{a \dot{b}}}
		\\ &
		+ \Cdot_{\dot{a} \dot{b}}^{\dot{\alpha} \dot{\beta}} \epsilon_{\dot{a} \dot{b}} \epsilon^{\dot{\alpha} \dot{\beta}} \ket*{\theta_{a \dot{\alpha}} \theta_{b \dot{\beta}}}
		+ \Cundot_{a b}^{\alpha \beta} \epsilon_{a b} \epsilon^{\alpha \beta} \ket*{\eta_{\alpha \dot{a}} \eta_{\beta \dot{b}}}
	\\
	\dsT \ket*{Z_{\alpha \dot{\alpha}} Z_{\beta \dot{\beta}}} = {}&
		- 2 \Aalgebraic \ket*{Z_{\alpha \dot{\alpha}} Z_{\beta \dot{\beta}}}
		+ (-\Balgebraic + \Walgebraic \epsilon_{\dot{\alpha} \dot{\beta}}) \ket*{Z_{\alpha \dot{\beta}} Z_{\beta \dot{\alpha}}}
		+ (-\Balgebraic - \Walgebraic \epsilon_{\alpha \beta}) \ket*{Z_{\beta \dot{\alpha}} Z_{\alpha \dot{\beta}}}
		\\ &
		- \CbarDot_{\dot{\alpha} \dot{\beta}}^{\dot{a} \dot{b}} \epsilon_{\dot{\alpha} \dot{\beta}} \epsilon^{\dot{a} \dot{b}} \ket*{\eta_{\alpha \dot{a}} \eta_{\beta \dot{b}}}
		- \CbarUndot_{\alpha \beta}^{a b} \epsilon_{\alpha \beta} \epsilon^{a b} \ket*{\theta_{a \dot{\alpha}} \theta_{b \dot{\beta}}}
	\\
	\dsT \ket*{Y_{a \dot{a}} Z_{\alpha \dot{\alpha}}} = {}&
		+ 2 \Galgebraic \ket*{Y_{a \dot{a}} Z_{\alpha \dot{\alpha}}}
		+ \Hundot_{a \alpha}^{\alpha a} \ket*{\eta_{\alpha \dot{a}} \theta_{a \dot{\alpha}}}
		- \Hdot_{\dot{a} \dot{\alpha}}^{\dot{\alpha} \dot{a}} \ket*{\theta_{a \dot{\alpha}} \eta_{\alpha \dot{a}}}
	\\
	\dsT \ket*{Z_{\alpha \dot{\alpha}} Y_{a \dot{a}}} = {}&
		- 2 \Galgebraic \ket*{Z_{\alpha \dot{\alpha}} Y_{a \dot{a}}}
		+ \HbarDot_{\dot{\alpha} \dot{a}}^{\dot{a} \dot{\alpha}} \ket*{\eta_{\alpha \dot{a}} \theta_{a \dot{\alpha}}}
		- \HbarUndot_{\alpha a}^{a \alpha} \ket*{\theta_{a \dot{\alpha}} \eta_{\alpha \dot{a}}}
\myintertext{Fermion-Fermion}
	\dsT \ket*{\theta_{a \dot{\alpha}} \theta_{b \dot{\beta}}} = {}&
		+ \CbarDot_{\dot{\alpha} \dot{\beta}}^{\dot{a} \dot{b}} \epsilon_{\dot{\alpha} \dot{\beta}} \epsilon^{\dot{a} \dot{b}} \ket*{Y_{a \dot{a}} Y_{b \dot{b}}}
		- \Cundot_{a b}^{\alpha \beta} \epsilon_{a b} \epsilon^{\alpha \beta} \ket*{Z_{\alpha \dot{\alpha}} Z_{\beta \dot{\beta}}}
	\\
	\dsT \ket*{\eta_{\alpha \dot{a}} \eta_{\beta \dot{b}}} = {}&
		- \Cdot_{\dot{a} \dot{b}}^{\dot{\alpha} \dot{\beta}} \epsilon_{\dot{a} \dot{b}} \epsilon^{\dot{\alpha} \dot{\beta}} \ket*{Z_{\alpha \dot{\alpha}} Z_{\beta \dot{\beta}}}
		+ \CbarUndot_{\alpha \beta}^{a b} \epsilon_{\alpha \beta} \epsilon^{a b} \ket*{Y_{a \dot{a}} Y_{b \dot{b}}}
	\\
	\dsT \ket*{\theta_{a \dot{\alpha}} \eta_{\beta \dot{b}}} = {}&
		- \HbarDot_{\dot{\alpha} \dot{b}}^{\dot{b} \dot{\alpha}} \ket*{Y_{a \dot{b}} Z_{\beta \dot{\alpha}}}
		- \Hundot_{a \beta}^{\beta a} \ket*{Z_{\beta \dot{\alpha}} Y_{a \dot{b}}}
	\\
	\dsT \ket*{\eta_{\alpha \dot{a}} \theta_{b \dot{\beta}}} = {}&
		+ \Hdot_{\dot{a} \dot{\beta}}^{\dot{\beta} \dot{a}} \ket*{Z_{\alpha \dot{\beta}} Y_{b \dot{a}}}
		+ \HbarUndot_{\alpha b}^{b \alpha} \ket*{Y_{b \dot{a}} Z_{\alpha \dot{\beta}}}
\myintertext{Boson-Fermion}
	\dsT \ket*{Y_{a \dot{a}} \theta_{b \dot{\beta}}} = {}&
		(\Aalgebraic + \Galgebraic) \ket*{Y_{a \dot{a}} \theta_{b \dot{\beta}}}
		+ (\Balgebraic - \Walgebraic \epsilon_{a b}) \ket*{Y_{b \dot{a}} \theta_{a \dot{\beta}}}
		\\ &
		+ \Hdot_{\dot{a} \dot{\beta}}^{\dot{\beta} \dot{a}} \ket*{\theta_{a \dot{\beta}} Y_{b \dot{a}}}
		+ \Cundot_{a b}^{\alpha \beta} \epsilon_{a b} \epsilon^{\alpha \beta} \ket*{\eta_{\alpha \dot{a}} Z_{\beta \dot{\beta}}}
	\\
	\dsT \ket*{Y_{a \dot{a}} \eta_{\beta \dot{b}}} = {}&
		(\Aalgebraic + \Galgebraic) \ket*{Y_{a \dot{a}} \eta_{\beta \dot{b}}}
		+ (\Balgebraic + \Walgebraic \epsilon_{\dot{a} \dot{b}}) \ket*{Y_{a \dot{b}} \eta_{\beta \dot{a}}}
		\\ &
		+ \Hundot_{a \beta}^{\beta a} \ket*{\eta_{\beta \dot{a}} Y_{a \dot{b}}}
		- \Cdot_{\dot{a} \dot{b}}^{\dot{\alpha} \dot{\beta}} \epsilon_{\dot{a} \dot{b}} \epsilon^{\dot{\alpha} \dot{\beta}} \ket*{\theta_{a \dot{\alpha}} Z_{\beta \dot{\beta}}}
	\\
	\dsT \ket*{\theta_{a \dot{\alpha}} Y_{b \dot{b}}} = {}&
		(\Aalgebraic - \Galgebraic) \ket*{\theta_{a \dot{\alpha}} Y_{b \dot{b}}}
		+ (\Balgebraic - \Walgebraic \epsilon_{a b}) \ket*{\theta_{b \dot{\alpha}} Y_{a \dot{b}}}
		\\ &
		+ \HbarDot_{\dot{\alpha} \dot{b}}^{\dot{b} \dot{\alpha}} \ket*{Y_{a \dot{b}} \theta_{b \dot{\alpha}}}
		- \Cundot_{a b}^{\alpha \beta} \epsilon_{a b} \epsilon^{\alpha \beta} \ket*{Z_{\alpha \dot{\alpha}} \eta_{\beta \dot{b}}}
	\\
	\dsT \ket*{\eta_{\alpha \dot{a}} Y_{b \dot{b}}} = {}&
		(\Aalgebraic - \Galgebraic) \ket*{\eta_{\alpha \dot{a}} Y_{b \dot{b}}}
		+ (\Balgebraic + \Walgebraic \epsilon_{\dot{a} \dot{b}}) \ket*{\eta_{\alpha \dot{b}} Y_{b \dot{a}}}
		\\ &
		+ \HbarUndot_{\alpha b}^{b \alpha} \ket*{Y_{b \dot{a}} \eta_{\alpha \dot{b}}}
		+ \Cdot_{\dot{a} \dot{b}}^{\dot{\alpha} \dot{\beta}} \epsilon_{\dot{a} \dot{b}} \epsilon^{\dot{\alpha} \dot{\beta}} \ket*{Z_{\alpha \dot{\alpha}} \theta_{b \dot{\beta}}}
\displaybreak[3]
\\[15pt]
	\dsT \ket*{Z_{\alpha \dot{\alpha}} \theta_{b \dot{\beta}}} = {}&
		- (\Aalgebraic + \Galgebraic) \ket*{Z_{\alpha \dot{\alpha}} \theta_{b \dot{\beta}}}
		+ (-\Balgebraic + \Walgebraic \epsilon_{\dot{\alpha} \dot{\beta}}) \ket*{Z_{\alpha \dot{\beta}} \theta_{b \dot{\alpha}}}
		\\ &
		- \HbarUndot_{\alpha b}^{b \alpha} \ket*{\theta_{b \dot{\alpha}} Z_{\alpha \dot{\beta}}}
		+ \CbarDot_{\dot{\alpha} \dot{\beta}}^{\dot{a} \dot{b}} \epsilon_{\dot{\alpha} \dot{\beta}} \epsilon^{\dot{a} \dot{b}} \ket*{\eta_{\alpha \dot{a}} Y_{b \dot{b}}}
	\\
	\dsT \ket*{Z_{\alpha \dot{\alpha}} \eta_{\beta \dot{b}}} = {}&
		- (\Aalgebraic + \Galgebraic) \ket*{Z_{\alpha \dot{\alpha}} \eta_{\beta \dot{b}}}
		+ (-\Balgebraic - \Walgebraic \epsilon_{\alpha \beta}) \ket*{Z_{\beta \dot{\alpha}} \eta_{\alpha \dot{b}}}
		\\ &
		- \HbarDot_{\dot{\alpha} \dot{b}}^{\dot{b} \dot{\alpha}} \ket*{\eta_{\alpha \dot{b}} Z_{\beta \dot{\alpha}}}
		- \CbarUndot_{\alpha \beta}^{a b} \epsilon_{\alpha \beta} \epsilon^{a b} \ket*{\theta_{a \dot{\alpha}} Y_{b \dot{b}}}
	\\
	\dsT \ket*{\theta_{a \dot{\alpha}} Z_{\beta \dot{\beta}}} = {}&
		- (\Aalgebraic - \Galgebraic) \ket*{\theta_{a \dot{\alpha}} Z_{\beta \dot{\beta}}}
		+ (-\Balgebraic + \Walgebraic \epsilon_{\dot{\alpha} \dot{\beta}}) \ket*{\theta_{a \dot{\beta}} Z_{\beta \dot{\alpha}}}
		\\ &
		- \Hundot_{a \beta}^{\beta a} \ket*{Z_{\beta \dot{\alpha}} \theta_{a \dot{\beta}}}
		- \CbarDot_{\dot{\alpha} \dot{\beta}}^{\dot{a} \dot{b}} \epsilon_{\dot{\alpha} \dot{\beta}} \epsilon^{\dot{a} \dot{b}} \ket*{Y_{a \dot{a}} \eta_{\beta \dot{b}}}
	\\
	\dsT \ket*{\eta_{\alpha \dot{a}} Z_{\beta \dot{\beta}}} = {}&
		- (\Aalgebraic - \Galgebraic) \ket*{\eta_{\alpha \dot{a}} Z_{\beta \dot{\beta}}}
		+ (-\Balgebraic - \Walgebraic \epsilon_{\alpha \beta}) \ket*{\eta_{\beta \dot{a}} Z_{\alpha \dot{\beta}}}
		\\ &
		- \Hdot_{\dot{a} \dot{\beta}}^{\dot{\beta} \dot{a}} \ket*{Z_{\alpha \dot{\beta}} \eta_{\beta \dot{a}}}
		+ \CbarUndot_{\alpha \beta}^{a b} \epsilon_{\alpha \beta} \epsilon^{a b} \ket*{Y_{a \dot{a}} \theta_{b \dot{\beta}}}
\end{align*}
\endgroup
Because we work only up to quadratic order in fermions, we were not able to determine the expressions for four-fermion processes.
The coefficients used above are defined as
\newcommand{\Ccommon}{\Calgebraic_0}
\newcommand{\Hcommon}{\Halgebraic_0}
\begin{equation}
	\label{T-matrix-coefficients}
	\begin{aligned}
		\Aalgebraic & = \frac{1}{4} \frac{(p_1 - p_2)^2 + \kappa^2 (\omega_1 - \omega_2)^2}{p_1 \omega_2 - p_2 \omega_1}
		\mcomma \\
		\Balgebraic & = \frac{p_1 p_2 + \kappa^2 \omega_1 \omega_2}{p_1 \omega_2 - p_2 \omega_1}
		\mcomma \\
		\Galgebraic & = - \qty(1+\kappa^2) \frac{1}{4} \frac{\omega_1^2 - \omega_2^2}{p_1 \omega_2 - p_2 \omega_1}
		\mcomma \\
		\Walgebraic & = \I \kappa
		\mcomma \\
		\Ccommon & =
			- \qty(1 + \kappa^2) \sqrt{p_1^2 + \kappa^2} \sqrt{p_2^2 + \kappa^2}
			\frac{\sinh(\frac{1}{2} (\arsinh \frac{p_1}{\sqrt{1+\kappa^2}} - \arsinh \frac{p_2}{\sqrt{1+\kappa^2}}))}{p_1 \omega_2 - p_2 \omega_1}
		\mcomma \\
		\Hcommon & =
			+ \qty(1 + \kappa^2) \sqrt{p_1^2 + \kappa^2} \sqrt{p_2^2 + \kappa^2}
			\frac{\cosh(\frac{1}{2} (\arsinh \frac{p_1}{\sqrt{1+\kappa^2}} - \arsinh \frac{p_2}{\sqrt{1+\kappa^2}}))}{p_1 \omega_2 - p_2 \omega_1}
		\mcomma \\[10pt]
		\phantom{\Calgebraic_{12}^{34}(\kappa)}
		& \begin{alignedat}{3}
			\mathllap{\Calgebraic_{12}^{34}(\kappa)} & = \frac{p_1 - \I \kappa \omega_1}{p_2 - \I \kappa \omega_2} \frac{p_2 + \I \kappa}{p_1 - \I \kappa} \Ccommon
			\mcomma \qquad &
			\Calgebraic_{12}^{43}(\kappa) & = \Ccommon
			\mcomma \qquad &
			\CbarAlgebraic_{\alpha \beta}^{a b}(\kappa) & = (\Calgebraic_{a b}^{\alpha \beta}(\kappa))^*
			\mcomma \\
			\mathllap{\Calgebraic_{21}^{43}(\kappa)} & = \frac{p_2 - \I \kappa \omega_2}{p_1 - \I \kappa \omega_1} \frac{p_1 + \I \kappa}{p_2 - \I \kappa} \Ccommon
			\mcomma \qquad &
			\Calgebraic_{21}^{34}(\kappa) & = \Ccommon
			\mcomma \\[10pt]
			\mathllap{\Halgebraic_{13}^{31}(\kappa)} & = \frac{p_2 + \I \kappa \omega_2}{p_1 + \I \kappa \omega_1} \frac{p_1 + \I \kappa}{p_2 + \I \kappa} \Hcommon
			\mcomma \qquad &
			\Halgebraic_{14}^{41}(\kappa) & = \Hcommon
			\mcomma \qquad &
			\HbarAlgebraic_{\alpha b}^{a \beta}(\kappa) & = (\Halgebraic_{a \beta}^{\alpha b}(\kappa))^*
			\mcomma \\
			\mathllap{\Halgebraic_{24}^{42}(\kappa)} & = \frac{p_2 - \I \kappa \omega_2}{p_1 - \I \kappa \omega_1} \frac{p_1 + \I \kappa}{p_2 + \I \kappa} \Hcommon
			\mcomma \qquad &
			\Halgebraic_{23}^{32}(\kappa) & = \Hcommon
			\mcomma
		\end{alignedat}
		\\[12pt]
		& \begin{alignedat}{4}
			\mathllap{\Cundot} & = \Calgebraic(-\kappa)
			\mcomma \qquad &
			\CbarUndot & = \CbarAlgebraic(-\kappa)
			\mcomma \qquad &
			\Hundot & = \Halgebraic(-\kappa)
			\mcomma \qquad &
			\HbarUndot & = \HbarAlgebraic(-\kappa)
			\mdot
		\end{alignedat}
	\end{aligned}
\end{equation}
$\dsT$ satisfies the classical Yang-Baxter equation%
\footnote{
	\label{definition-graded-embeddings-T}%
	The $\dsT_{ij}$ denote the graded embeddings of $\dsT$ into the product of three spaces, i.e.\ using the graded permutation operator $P^\mup{g}_{ij}$ (defined, for example, in eq.\ (3.8) of \cite{Arutyunov:2009ga}).
	Explicitly this gives
	\begin{equation*}
		\dsT_{12} = \dsT \otimes \identity
		\mcomma \qquad
		\dsT_{13} = P^\mup{g}_{23} \dsT_{12} P^\mup{g}_{23}
		\mcomma \qquad
		\dsT_{23} = P^\mup{g}_{12} P^\mup{g}_{13} \dsT_{12} P^\mup{g}_{13} P^\mup{g}_{12} = \identity \otimes \dsT
		\mdot
	\end{equation*}
}
\begin{equation}
	\commutator{\dsT_{23}}{\dsT_{13}}
	+ \commutator{\dsT_{23}}{\dsT_{12}}
	+ \commutator{\dsT_{13}}{\dsT_{12}}
	= 0
\end{equation}
up to terms that could not be checked because they involve four-fermion expressions.
Additionally, in the undeformed limit $\kappa \to 0$ it matches the result of \cite{Klose:2006zd} up to a sign factor $\sgn(p_1) \cdot \sgn(p_2)$ in fermionic amplitudes; see also the discussion in \cref{explanation-phase-f}.

\subsection{Factorization}
\label{T-matrix-factorization}
Our tree-level result matches a T matrix written in the factorized form
\begin{equation}
	\label{factorized-form-of-T}
	\begin{gathered}
		\dsT = \scT(-\kappa) \otimes \identity + \identity \otimes \scT(\kappa)
		\mcomma
		\\
		\dsT^{P \dot{P} Q \dot{Q}}_{M \dot{M} N \dot{N}} =
			(-1)^{\epsilon_{\dot{M}} (\epsilon_N + \epsilon_Q)}
			\scT^{PQ}_{MN}(-\kappa) \delta^{\dot{P}}_{\dot{M}} \delta^{\dot{Q}}_{\dot{N}}
			+
			(-1)^{\epsilon_Q (\epsilon_{\dot{M}} + \epsilon_{\dot{P}})}
			\delta^P_M \delta^Q_N \scT^{\dot{P}\dot{Q}}_{\dot{M}\dot{N}}(\kappa)
			\mcomma
	\end{gathered}
\end{equation}
up to the four-fermion amplitudes that we did not compute.
Here the first and second factor of the tensor product acts respectively on the undotted or dotted indices.
$\epsilon_M$ describes the statistics of the index, i.e.\ it is zero for Latin indices (1 and 2) and one for Greek indices (3 and 4).
The matrix $\scT$ is the tree-level expansion of the exact fermionic $\mathfrak{su}_q(2|2)_\mup{c.e.}$ $S'$ matrix that will be derived in the next section, explicitly in subsection~\ref{exact-fermionic-S-matrix}.
The entries of $\scT(\kappa)$ are
\begin{equation}
	\label{definition-scT}
	\begin{aligned}
		& \scT_{ab}^{cd} = \Aalgebraic \delta_a^c \delta_b^d + (\Balgebraic + \Walgebraic \epsilon_{ab}) \delta_a^d \delta_b^c
		\mcomma \\
		& \scT_{\alpha \beta}^{\gamma \delta} = -\Aalgebraic \delta_\alpha^\gamma \delta_\beta^\delta +
			(-\Balgebraic + \Walgebraic \epsilon_{\alpha \beta}) \delta_\alpha^\delta \delta_\beta^\gamma
		\mcomma \\
		& \begin{alignedat}{2}
			\scT_{a \beta}^{c \delta} & = \Galgebraic \delta_a^c \delta_\beta^\delta
			\mcomma \qquad \quad &
			\scT_{\alpha b}^{\gamma d} & = -\Galgebraic \delta_\alpha^\gamma \delta_b^d
			\mcomma \\
			\scT_{a b}^{\gamma \delta} & = \Calgebraic_{a b}^{\gamma \delta} \epsilon_{a b} \epsilon^{\gamma \delta}
			\mcomma &
			\scT_{\alpha \beta}^{c d} & = \CbarAlgebraic_{\alpha \beta}^{c d} \epsilon_{\alpha \beta} \epsilon^{c d}
			\mcomma \\
			\scT_{a \beta}^{\gamma d} & = \Halgebraic_{a \beta}^{\gamma d} \delta_a^d \delta_\beta^\gamma
			\mcomma &
			\scT_{\alpha b}^{c \delta} & = \HbarAlgebraic_{\alpha b}^{c \delta} \delta_\alpha^\delta \delta_b^c
			\mcomma
		\end{alignedat}
	\end{aligned}
\end{equation}
with the coefficients listed in \eqref{T-matrix-coefficients}.
The sign flip of $\kappa$ in the first term of \cref{factorized-form-of-T} leaves the terms involving $\Aalgebraic$, $\Balgebraic$ and $\Galgebraic$ invariant and changes the $\Walgebraic$ term by a sign.
The $\Calgebraic$ and $\Halgebraic$ terms in turn transform in a non-trivial way.

In summary, we find a factorized T matrix that is structurally similar to the undeformed \cite{Klose:2006zd} and distinguished case \cite{Arutyunov:2015qva}, but differs in two major aspects.
Firstly, the coefficients $\Calgebraic$ and $\Halgebraic$ depend on the indices of their respective entries.
Secondly, in contrast to the results for the distinguished model presented in \cite{Arutyunov:2015qva}, the two factors in \cref{factorized-form-of-T} have opposite deformation parameters. We will come back to the origin of the relative sign flip on $\kappa$ in \cref{sec:comparison}, and the distinguished case in \cref{sec:distinguishedcase}. First we will determine the exact S matrix following purely from symmetry considerations.

\section{Exact S matrix}
\label{sec:exactSmatrix}

In this section we derive the exact $\mathfrak{su}_q(2|2)_\mup{c.e.}$ S matrix for the fermionic deformation. We exploit the fact that at the level of the complexified superalgebra $\mathfrak{sl}_q(2|2)_\mup{c.e.}$ the Hopf algebras constructed using respectively the distinguished and fully fermionic Dynkin diagram of $\mathfrak{sl}(2|2)$ have coproducts related by a twist. The S matrix associated to the fully fermionic Dynkin diagram can thus be obtained from the $\mathfrak{sl}_q(2|2)_\mup{c.e.}$ S matrix associated to the distinguished Dynkin diagram through twisting and upon imposing appropriate reality conditions.

\subsection{Hopf algebra}

Let us first recall the defining relations of the $\mathfrak{su}_q(2|2)$ superalgebra. For this we start by considering a Cartan-Weyl basis of the complexified $\mathfrak{sl}(2|2)$ superalgebra, formed by Cartan elements $\mathbb{H}_j$, positive roots $ \mathbb{E}_j$ and negative roots $ \mathbb{F}_j$, where the index $j=1,2,3$. The $q$-deformation is defined through the relations
\begin{equation}
\label{eq:qdef}
q^{\mathbb{H}_j} \mathbb{E}_k = q^{ A_{jk}}  \mathbb{E}_k q^{ \mathbb{H}_j} , \qquad q^{ \mathbb{H}_j} \mathbb{F}_k = q^{- A_{jk}}  \mathbb{F}_k q^{ \mathbb{H}_j} , \qquad \com{ \mathbb{E}_j}{ \mathbb{F}_k} = d_j \delta_{jk} \Q{\mathbb{H}_j} \mcomma
\end{equation}
with $\Q{x} = (q^x - q^{-x}) /(q-q^{-1})$ and $ A$ a symmetric Cartan matrix associated to $\mathfrak{sl}(2|2)$, obtained from the original unsymmetrized Cartan matrix $\hat A$ through $\hat{A}=D A$ with $D=\diag(d_1,d_2,d_3)$. A particularity of Lie superalgebras that sets them apart from ordinary Lie algebras is that they admit inequivalent Dynkin diagrams, depending on the number of bosonic simple roots in the chosen root system. Each Dynkin diagram is associated to a different Cartan matrix.  $\mathfrak{sl}(2|2)$ admits three inequivalent Dynkin diagrams. We will focus on two of them, the distinguished Dynkin diagram, which has the maximum number of bosonic simple roots (two), and the fermionic Dynkin diagram, where all the three simple roots are fermionic.

The relations \eqref{eq:qdef} are not enough to completely fix the $\mathfrak{sl}_q(2|2)$ superalgebra but need to be supplemented with standard and higher-order Serre relations. We do not write these conditions in their most general form here (independent on the choice of Dynkin diagram), but rather later when considering the distinguished and fermionic Dynkin diagram. Some of these constraints can be consistently dropped, giving rise to a centrally extended $\mathfrak{sl}_q(2|2)_\mup{c.e.}$ superalgebra.

There are several coproducts under which the quantum-deformed algebra acquires a coalgebra structure. Here we choose the one whose action on the Cartan elements, positive and negative simple roots is given by
\begin{equation}
\label{eq:coproduct}
	\begin{aligned}
		\Delta(\mathbb{H}_j) &= \mathbb{H}_j \otimes 1 + 1 \otimes \mathbb{H}_j \mcomma \\
		\Delta(\mathbb{E}_j) &= \mathbb{E}_j \otimes 1  + q^{-\mathbb{H}_j} \otimes \mathbb{E}_j \mcomma \\
		\Delta(\mathbb{F}_j) &= \mathbb{F}_j \otimes q^{\mathbb{H}_j}  + 1  \otimes \mathbb{F}_j \mdot
	\end{aligned}
\end{equation}
This coproduct satisfies the required conditions
\begin{equation}
(1 \otimes \Delta) \Delta = (\Delta \otimes 1) \Delta \mcomma \qquad (1 \otimes \varepsilon ) \Delta = 1 = (\varepsilon \otimes 1) \Delta \mcomma
\end{equation}
where $1$ denotes the identity and $\varepsilon \colon \mathfrak{sl}_q(2|2)_\mup{c.e.} \to \mathbb C$ is the counit with
\begin{equation}
\varepsilon(1)=1 \mcomma \qquad \varepsilon(\mathbb H_j)=\varepsilon(\mathbb E_j) = \varepsilon (\mathbb F_j) =0 \mdot
\end{equation}

Finally, to obtain a Hopf algebra we need to define an antipode map $\antipode \colon \mathfrak{sl}_q(2|2)_\mup{c.e.} \to \mathfrak{sl}_q(2|2)_\mup{c.e.}$ satisfying the compatibility condition
\begin{equation}
\label{eq:antipode}
\mu (\antipode \otimes 1) \Delta(X) = \mu (1 \otimes \antipode) \Delta(X) = \eta \varepsilon (X) \mcomma
\end{equation}
where $\mu \colon \mathfrak{sl}_q(2|2)_\mup{c.e.} \otimes \mathfrak{sl}_q(2|2)_\mup{c.e.} \to \mathfrak{sl}_q(2|2)_\mup{c.e.} $ denotes the product, $\mu (X \otimes Y) = XY$.

Let us already mention that in order to obtain a non-trivial S matrix one needs to introduce the braiding into the coproduct \eqref{eq:coproduct}. The coproduct of Cartan elements and bosonic simple roots remains unchanged, but the coproduct of fermionic simple roots needs to be adapted. We postpone the explicit expression of the coproduct with braiding to when we consider specific Dynkin diagrams.

\paragraph{Distinguished Dynkin diagram.}
The distinguished Cartan matrix corresponds to choosing a root system with the maximum number of bosonic simple roots. In the case of $\mathfrak{sl}(2|2)$, this corresponds to two bosonic simple roots and one fermionic simple root. This is the Dynkin diagram chosen in \cite{Beisert:2008tw}, and we review the main characteristics of the corresponding Hopf algebra here.

The unsymmetrized and symmetrized distinguished Cartan matrices are
\begin{equation}
\label{eq:Cmatrix_oxo}
\hat A=\begin{pmatrix}
+2 & -1 & 0 \\+1 & 0 & -1 \\ 0 & -1 & +2
\end{pmatrix} , \qquad
A= \begin{pmatrix}
+2 & -1 & 0 \\-1 & 0 & +1 \\ 0 & +1 & -2
\end{pmatrix} , \qquad D=\diag(+1,-1,-1) \mdot
\end{equation}
The standard Serre relations are
\begin{align}
0&= \mathbb{E}_1 \mathbb{E}_1 \mathbb{E}_2 -(q+q^{-1}) \mathbb{E}_1 \mathbb{E}_2 \mathbb{E}_1 + \mathbb{E}_2 \mathbb{E}_1 \mathbb{E}_1= \mathbb{E}_3 \mathbb{E}_3 \mathbb{E}_2 -(q+q^{-1}) \mathbb{E}_3 \mathbb{E}_2 \mathbb{E}_3 + \mathbb{E}_2 \mathbb{E}_3 \mathbb{E}_2 \mcomma\nonumber\\
&= \mathbb{F}_1 \mathbb{F}_1 \mathbb{F}_2 -(q+q^{-1}) \mathbb{F}_1 \mathbb{F}_2 \mathbb{F}_1 + \mathbb{F}_2 \mathbb{F}_1 \mathbb{F}_1= \mathbb{F}_3 \mathbb{F}_3 \mathbb{F}_2 -(q+q^{-1}) \mathbb{F}_3 \mathbb{F}_2 \mathbb{F}_3 + \mathbb{F}_2 \mathbb{F}_3 \mathbb{F}_2 \mcomma\nonumber\\
&=  \com{\mathbb{E}_1}{\mathbb{E}_3}= \com{\mathbb{F}_1}{\mathbb{F}_3}=\mathbb{E}_2 \mathbb{E}_2 = \mathbb{F}_2 \mathbb{F}_2 \mcomma
\end{align}
and the higher order Serre relations take the form $\mathbb{P}=0$ and $\mathbb{K}=0$ with
\begin{equation}
\label{eq:Serre_oxo_2}
\begin{aligned}
\mathbb{P}&= \mathbb{E}_1 \mathbb{E}_2 \mathbb{E}_3 \mathbb{E}_2 + \mathbb{E}_2 \mathbb{E}_3 \mathbb{E}_2 \mathbb{E}_1 + \mathbb{E}_3 \mathbb{E}_2 \mathbb{E}_1 \mathbb{E}_2 + \mathbb{E}_2 \mathbb{E}_1 \mathbb{E}_2 \mathbb{E}_3 - (q+q^{-1}) \mathbb{E}_2 \mathbb{E}_1 \mathbb{E}_3 \mathbb{E}_2 \mcomma \\
\mathbb{K}&= \mathbb{F}_1 \mathbb{F}_2 \mathbb{F}_3 \mathbb{F}_2 + \mathbb{F}_2 \mathbb{F}_3 \mathbb{F}_2 \mathbb{F}_1 + \mathbb{F}_3 \mathbb{F}_2 \mathbb{F}_1 \mathbb{F}_2 + \mathbb{F}_2 \mathbb{F}_1 \mathbb{F}_2 \mathbb{F}_3 - (q+q^{-1}) \mathbb{F}_2 \mathbb{F}_1 \mathbb{F}_3 \mathbb{F}_2 \mdot
\end{aligned} \end{equation}
The Cartan matrix has non-maximal rank 2 and there is thus a central element, given by $\mathbb{C} = -\mathbb{H}_2 - \frac{1}{2} ( \mathbb{H}_1 + \mathbb{H}_3)$. In fact, it can be shown that the higher-order Serre relations \eqref{eq:Serre_oxo_2} can be consistently dropped, in which case also $\mathbb{P}$ and $\mathbb{K}$ become central elements and one obtains the triply centrally extended algebra $\mathfrak{sl}_q(2|2) \ltimes \mathbb{R}^2$.

The coproduct (including the braiding factor $\mathbb{U}$) of the Cartan elements and simple roots is%
\footnote{In contrast to \cite{Beisert:2008tw} we choose to include the braiding in a ``symmetric'' way. This choice is more suited for implementing the twist.}
\begin{equation}
\label{eq:coproduct1_oxo}
\begin{aligned}
\Delta(\mathbb{H}_j) &= \mathbb{H}_j \otimes 1 + 1 \otimes \mathbb{H}_j \mcomma \\
\Delta(\mathbb{E}_j) &= \left\{
\begin{aligned}
&\mathbb{E}_{j} \otimes 1  + q^{-\mathbb{H}_{j}} \otimes \mathbb{E}_{j} &\qquad &j=1,3 \mcomma \\
&\mathbb{E}_j \otimes \mathbb{U}^{-1/2}  + q^{-\mathbb{H}_j} \mathbb{U}^{1/2} \otimes \mathbb{E}_j&\qquad &j=2 \mcomma
\end{aligned}
\right. \\
\Delta(\mathbb{F}_j) &= \left\{
\begin{aligned}
&\mathbb{F}_j \otimes q^{\mathbb{H}_j}  + 1  \otimes \mathbb{F}_j &\qquad &\,\,\,j=1,3 \mcomma \\
&\mathbb{F}_j \otimes q^{\mathbb{H}_j} \mathbb{U}^{1/2}  + \mathbb{U}^{-1/2}  \otimes \mathbb{F}_j &\qquad &\,\,\,j=2 \mdot
\end{aligned}
\right.
\end{aligned}
\end{equation}
This in turn fixes the coproduct of the three central elements to be
\begin{equation} \begin{aligned}
\Delta(\mathbb{C}) &= \mathbb{C} \otimes 1 + 1 \otimes \mathbb{C} \mcomma \\
\Delta(\mathbb{P}) &= \mathbb{P} \otimes \mathbb{U}^{-1} + q^{2 \mathbb{C}} \mathbb{U} \otimes \mathbb{P} \mcomma \\
\Delta(\mathbb{K}) &= \mathbb{K} \otimes q^{-2 \mathbb{C}} \mathbb{U} + \mathbb{U}^{-1} \otimes \mathbb{K} \mdot
\end{aligned}
\end{equation}
The S matrix should satisfy%
\footnote{For an operator $\mathcal O$ mapping to the tensor product of two spaces, we define the ``opposite'' operator $\mathcal O^\mup{op}=P^\mup{g} \mathcal O$, where $P^\mup{g}$ is the graded permutation operator. In particular, for the coproduct we have $\Delta^\mup{op} (X)= P^\mup{g}(\Delta(X))$.}
\begin{equation}
\label{eq:S_oxo}
\Delta^\mup{op} (X) S = S \Delta(X) \mcomma \qquad \forall X \in \mathfrak{sl}_q(2|2)_\mup{c.e.} \mdot
\end{equation}
 In particular if $X$ is central this implies $\Delta^\mup{op}(X)= \Delta(X)$. While this is immediately satisfied for $\mathbb{C}$, imposing it for $\mathbb{P}$ and $\mathbb{K}$ partially fixes them to be
\begin{equation}
\label{eq:PK_oxo}
\mathbb{P}= \alpha \beta \, \mathbb{U}^{-1} \left( 1- q^{2 \mathbb{C}} \mathbb{U}^{2} \right) \mcomma \qquad \mathbb{K}=  \alpha^{-1} \beta \, \mathbb{U} \left( q^{-2 \mathbb{C}} - \mathbb{U}^{-2} \right) \mcomma
\end{equation}
where $\alpha$ and $\beta$ are yet undetermined complex numbers.

Finally, the antipode map satisfying \eqref{eq:antipode} is given by $\antipode(1)=1$ and
\begin{equation}
\label{eq:antipode_oxo}
\begin{aligned}
\antipode(\mathbb H_j)&= - \mathbb H_j \mcomma &\qquad \antipode (\mathbb E_j) &= - q^{\mathbb H_j} \mathbb E_j \mcomma &\qquad \antipode (\mathbb F_j) &= - \mathbb F_j q^{- \mathbb H_j} \mcomma \\
\antipode(\mathbb C)&= - \mathbb C \mcomma &\qquad \antipode (\mathbb P) &= - q^{-2 \mathbb C} \mathbb P \mcomma &\qquad \antipode (\mathbb K) &= - \mathbb K q^{2 \mathbb C} \mdot
\end{aligned}
\end{equation}

\paragraph{Fermionic Dynkin diagram.}
For the fermionic Dynkin diagram on the other hand all the three simple roots are fermionic. The non-symmetrized and symmetrized Cartan matrices are  (we use primes to denote quantities related to the fermionic Dynkin diagram)
\begin{equation}
\label{eq:Cmatrix_xxx}
	\hat A'=\begin{pmatrix}
		0 & -1 & 0 \\+1 & 0 & -1 \\ 0 & +1 & 0
	\end{pmatrix}
	, \quad
	A'= \begin{pmatrix}
		0 & +1 & 0 \\+1 & 0 & -1 \\ 0 & -1 & 0
	\end{pmatrix}
	, \quad
	D'=\diag(-1,+1,-1)
	\mdot
\end{equation}
The standard Serre relations are
\begin{equation}
\label{eq:Serre_xxx_1}
0=\mathbb{E}_1' \mathbb{E}_1' = \mathbb{F}_1' \mathbb{F}_1' = \mathbb{E}_2' \mathbb{E}_2' = \mathbb{F}_2' \mathbb{F}_2' = \mathbb{E}_3' \mathbb{E}_3' = \mathbb{F}_3' \mathbb{F}_3' \mcomma
\end{equation}
together with $\mathbb{P}'=0$ and $\mathbb{K}'=0$, where
\begin{equation}
\label{eq:Serre_xxx_2}
\mathbb{P}'= \com{\mathbb{E}_1'}{\mathbb{E}_3'} \mcomma \qquad \mathbb{K}'= \com{\mathbb{F}_1'}{\mathbb{F}_3'} \mdot
\end{equation}
The higher-order Serre relations are automatically satisfied. The reason why we have separated the standard Serre relations into \eqref{eq:Serre_xxx_1} and \eqref{eq:Serre_xxx_2} is that the second constraints can be consistently dropped, leading to a triply centrally extended $\mathfrak{sl}_q(2|2)$ superalgebra with central elements $\mathbb{P}', \mathbb{K}'$ and
\begin{equation}
\mathbb{C}'=-\frac{1}{2} (\mathbb{H}'_1+\mathbb{H}'_3) \mdot
\end{equation}

The coproduct (including the braiding) is
\begin{equation}
\begin{aligned}
\label{eq:coproduct1_xxx}
\Delta'(\mathbb{H}_j') &= \mathbb{H}_j' \otimes 1 + 1 \otimes \mathbb{H}_j' \mcomma \\
\Delta'(\mathbb{E}_j') &= \left\{ \begin{aligned} & \mathbb{E}_{j}' \otimes \mathbb{U}^{-1/2}  + q^{-\mathbb{H}_{j}'} \mathbb{U}^{1/2} \otimes \mathbb{E}'_{j} &\qquad &j=1,3 \mcomma \\
&\mathbb{E}_j' \otimes \mathbb{U}^{1/2}  + q^{-\mathbb{H}_j'} \mathbb{U}^{-1/2} \otimes \mathbb{E}'_j &\qquad &j=2 \mcomma \end{aligned} \right. \\
\Delta'(\mathbb{F}_j) &= \left\{ \begin{aligned}
&\mathbb{F}_j' \otimes q^{\mathbb{H}_j'} \mathbb{U}^{1/2}  + \mathbb{U}^{-1/2}  \otimes \mathbb{F}'_j &\qquad \, \, \, &j=1,3 \mcomma \\
&\mathbb{F}_j' \otimes q^{\mathbb{H}_j'} \mathbb{U}^{-1/2}  + \mathbb{U}^{1/2}  \otimes \mathbb{F}'_j &\qquad \, \, \,&j=2 \mdot
\end{aligned} \right.
\end{aligned}
\end{equation}
For the three central elements we obtain
\begin{equation} \begin{aligned}
\Delta'(\mathbb{C}') &= \mathbb{C}' \otimes 1 + 1 \otimes \mathbb{C}' \mcomma \\
\Delta'(\mathbb{P}') &= \mathbb{P}' \otimes \mathbb{U}^{-1} + q^{2 \mathbb{C}'} \mathbb{U} \otimes \mathbb{P}' \mcomma \\
\Delta'(\mathbb{K}') &= \mathbb{K}' \otimes q^{-2 \mathbb{C}'} \mathbb{U} + \mathbb{U}^{-1} \otimes \mathbb{K}' \mcomma
\end{aligned}
\end{equation}
and $\mathbb P'$, $\mathbb K'$ obey analogous relations to \eqref{eq:PK_oxo}.
The antipode map $\antipode'$ is given by equations similar to \eqref{eq:antipode_oxo}.

\paragraph{The twist.} As already mentioned, the $q$-deformed superalgebra generated by the fer\-mi\-o\-nic Cartan matrix is isomorphic to the $q$-deformed superalgebra generated by the distinguished Cartan matrix, while the coproducts are related by a twist. This remains true even after introducing the braiding. To see this, we define the two distinct algebras $\Alg{A}(\{\mathbb H, \mathbb E, \mathbb F\}, A)$ and  $\Alg{A'}(\{\mathbb H', \mathbb E', \mathbb F'\}, A')$, whose defining relations \eqref{eq:qdef} are dictated by the Cartan matrices $A$ of \eqref{eq:Cmatrix_oxo} and $A'$ of \eqref{eq:Cmatrix_xxx} respectively. The Lusztig transformation $\omega \colon \Alg{A'}(\{\mathbb H', \mathbb E', \mathbb F'\}, A') \to \Alg{A}(\{\mathbb H, \mathbb E, \mathbb F\}, A)$ defined by \cite{Khoroshkin:1994uj}\footnote{I.e.\ we can think of $\omega(X')$ as the representation of $X'$ in the algebra $\Alg{A}(\{\mathbb H, \mathbb E, \mathbb F\}, A)$.}
\begin{fleqn}[5pt]
\begin{equation}
	\label{eq:LusztigTransform}
	\newcommand{\columnspacing}{\qquad}
	\begin{alignedat}{3}
		\omega(\mathbb{H}_1') &= \mathbb{H}_1 + \mathbb{H}_2 \mcomma \columnspacing &
		\omega(\mathbb{E}_1') &= \mathbb{E}_1 \mathbb{E}_2 -q \mathbb{E}_2 \mathbb{E}_1 \mcomma \columnspacing &
		\omega(\mathbb{F}_1') &= \mathrlap{\mathbb{F}_2 \mathbb{F}_1 - q^{-1}  \mathbb{F}_1 \mathbb{F}_2 \mcomma}
		\\
		\omega(\mathbb{H}_2') &= -\mathbb{H}_2 \mcomma \columnspacing &
		\omega(\mathbb{E}_2') &= -\mathbb{F}_2 q^{\mathbb{H}_2} \mcomma \columnspacing &
		\omega(\mathbb{F}_2') &= -q^{-\mathbb{H}_2} \mathbb{E}_2 \mcomma
		\\
		\omega(\mathbb{H}_3') &= \mathbb{H}_2 + \mathbb{H}_3 \mcomma \columnspacing &
		\omega(\mathbb{E}_3') &= \mathbb{E}_3 \mathbb{E}_2 - q^{-1} \mathbb{E}_2 \mathbb{E}_3 \mcomma \columnspacing &
		\omega(\mathbb{F}_3') &= \mathbb{F}_2 \mathbb{F}_3 -q \mathbb{F}_3 \mathbb{F}_2 \mcomma
	\end{alignedat}
\end{equation}
\end{fleqn}
is such that%
\footnote{On the left-hand side of \eqref{eq:homomorphism} the bracket is defined with the Cartan matrix $A$ of \eqref{eq:Cmatrix_oxo}, while on the right-hand side the bracket is defined with the Cartan matrix $A'$ \eqref{eq:Cmatrix_xxx}.}
\begin{equation}
\label{eq:homomorphism}
\com{\omega(X')}{\omega(Y')} =\omega( \com{X'}{Y'}) \mcomma \qquad \forall X', Y' \in \Alg{A'}(\{\mathbb H', \mathbb E', \mathbb F'\}, A') \mcomma
\end{equation}
and thus defines a Lie algebra homomorphism. Moreover, under this map the central elements are transformed into one another
\begin{equation}
\omega(\mathbb{C}')=\mathbb{C} \mcomma \qquad \omega(\mathbb{P}')=\mathbb{P} \mcomma\qquad \omega(\mathbb{K}')= \mathbb{K} \mdot
\end{equation}
The coproducts on the other hand are related by a Drinfel'd twist
\begin{equation}
(\omega \otimes \omega) \Delta' (X') = \Ftwist^{-1} \Delta(\omega(X')) \Ftwist \mcomma \qquad \Ftwist= 1\otimes1 - (q-q^{-1}) \mathbb{U}^{1/2} \mathbb{F}_2 \otimes \mathbb{U}^{1/2} \mathbb{E}_2 \mcomma
\end{equation}
with $\Ftwist$ satisfying the cocycle condition
\begin{equation}
(\Ftwist^{-1} \otimes 1)(\Delta \otimes 1) \Ftwist^{-1} = (1 \otimes \Ftwist^{-1}) (1 \otimes \Delta) \Ftwist^{-1} \mdot
\end{equation}
Therefore, the S matrix for the fermionic Dynkin diagram is
\begin{equation}
\label{eq:Smatrix_xxx}
S' = \Ftwist^{-\mup{op}} S \Ftwist \mdot
\end{equation}

Let us mention that the antipode map is not preserved by the twist, in the sense that $\omega(\antipode'(X')) \neq \antipode(\omega (X'))$. We rather have that
\begin{equation}
\label{eq:antipodetwist}
\begin{aligned}
\omega (\antipode' (\mathbb H_j')) &=  \antipode (\omega(\mathbb H_j')) \mcomma & & \\
\omega (\antipode' (\mathbb E_1')) &= -q \antipode (\omega(\mathbb E_1')) \mcomma &\qquad  \omega (\antipode' (\mathbb F_1') &=- q^{-1} \antipode (\omega(\mathbb F_1')) \mcomma \\
\omega (\antipode' (\mathbb E_2')) &= \antipode (\omega(\mathbb E_2')) \mcomma &\qquad  \omega (\antipode' (\mathbb F_2') &= \antipode (\omega(\mathbb F_2')) \mcomma \\
\omega (\antipode' (\mathbb E_3')) &= -q^{-1} \antipode (\omega(\mathbb E_3')) \mcomma &\qquad  \omega (\antipode' (\mathbb F_3') &= -q \antipode (\omega(\mathbb F_3')) \mdot
\end{aligned}
\end{equation}

\paragraph{Reality conditions.} Until now we have worked with the complexified algebra $\mathfrak{sl}_q(2|2)$ and have not imposed any reality conditions to obtain $\mathfrak{su}_q(2|2)$. The S matrix $S'$ of \eqref{eq:Smatrix_xxx} is thus not a priori unitary. Imposing the reality conditions
\begin{equation}
\label{eq:rc_oxo}
\mathbb{H}_j^\dagger = \mathbb{H}_j \mcomma \qquad \mathbb{E}_j^{\dagger} = q^{-\mathbb{H}_j} \mathbb{F}_j \mcomma \qquad \mathbb{U}^{\dagger}=\mathbb{U}^{-1} \mcomma
\end{equation}
produces a unitary S matrix $S$ associated to the distinguished Dynkin diagram, but the sought after S matrix $S'$, associated to the fermionic Dynkin diagram, is not unitary due to the twist. Therefore, we need to adapt the reality conditions so that $S$ is not unitary but $S'$ is. In other words, instead of using the reality conditions \eqref{eq:rc_oxo} that are compatible with the coproduct \eqref{eq:coproduct1_oxo}, we choose reality conditions that are compatible with the coproduct \eqref{eq:coproduct1_xxx}. These are
\begin{equation}
{\mathbb{H}'_j}^\dagger = {\mathbb{H}'_j} \mcomma \qquad {\mathbb{E}'_j}^\dagger = q^{-\mathbb{H}'_j} \mathbb{F}'_j \mcomma \qquad \mathbb{U}^\dagger = \mathbb{U}^{-1} \mdot
\end{equation}
Imposing
\begin{equation}
\omega^\dagger(X') = \omega(X'^\dagger) \mcomma
\end{equation}
then gives rise to
\begin{equation}
\label{eq:rc_oxo_bis}
\mathbb{E}_1^\dagger = q^{-1} q^{-2 \mathbb{H}_2 - \mathbb{H}_1} \mathbb{F}_1 \mcomma \qquad \mathbb{E}_2^{\dagger} = q^{\mathbb{H}_2} \mathbb{F}_2 \mcomma \qquad \mathbb{E}_3^{\dagger} = q q^{-2 \mathbb{H}_2 - \mathbb{H}_3} \mathbb{F}_3 \mdot
\end{equation}
It thus follows that a way to obtain the exact $q$-deformed S matrix for the fermionic Dynkin diagram is to twist the $\mathfrak{su}_q(2|2)_\mup{c.e.}$ S matrix associated to the distinguished Dynkin diagram and impose the reality conditions \eqref{eq:rc_oxo_bis}.

\subsection{Fundamental S matrix}
The $q$-deformed S matrix based on the distinguished Dynkin diagram of $\mathfrak{sl}_q(2|2)$ has been derived by Beisert and Koroteev in \cite{Beisert:2008tw}. For completeness we review the construction here, with some slight changes. In particular, we use the symmetric braiding of \eqref{eq:coproduct1_oxo} and work with the shifted and rescaled variables of \cite{Beisert:2011wq}.

\paragraph{Fundamental representation.}
The fundamental representation of the centrally extended $\mathfrak{sl}_q(2|2)$ superalgebra is spanned by four states ${\ket{\phi_1}, \ket{\phi_2}, \ket{\psi_3}, \ket{\psi_4}}$, with $\ket{\phi_a}$ bosonic and $\ket{\psi_\alpha}$ fermionic, obeying
\begin{equation} \begin{aligned}
\mathbb{H}_1 \ket{\phi_1} &= - \ket{\phi_1} \mcomma &\qquad \mathbb{H}_2 \ket{\phi_1} &= - \left( C-1/2\right) \ket{\phi_1} \mcomma  & \mathbb{H}_3 \ket{\phi_1} &=0 \mcomma \\
\mathbb{H}_1 \ket{\phi_2} &= + \ket{\phi_2} \mcomma &\qquad \mathbb{H}_2 \ket{\phi_2} &= - \left( C+1/2\right) \ket{\phi_2} \mcomma& \mathbb{H}_3 \ket{\phi_2} &= 0 \mcomma\\
\mathbb{H}_1 \ket{\psi_4} &= 0 \mcomma&  \mathbb{H}_2 \ket{\psi_4} &= - \left( C+1/2\right) \ket{\psi_4} \mcomma   &\qquad \mathbb{H}_3 \ket{\psi_4} &= + \ket{\psi_4} \mcomma \\
\mathbb{H}_1 \ket{\psi_3} &= 0 \mcomma& \mathbb{H}_2 \ket{\psi_3} &= - \left( C-1/2\right) \ket{\psi_3} \mcomma&\qquad \mathbb{H}_3 \ket{\psi_3} &= - \ket{\psi_3} \mdot
\end{aligned} \end{equation}
where $C$ is the central charge for the fundamental representation. The action of the simple roots is given by
\begin{equation}
\label{eq:concreterootaction}
 \begin{aligned}
\mathbb{E}_1 \ket{\phi_1} &= \aalt \ket{\phi_2} \mcomma  & \qquad  \mathbb{F}_2 \ket{\phi_1} &= c \ket{\psi_3} \mcomma  \\
\mathbb{E}_2 \ket{\phi_2} &= a \ket{\psi_4} \mcomma & \qquad  \mathbb{F}_1 \ket{\phi_2} &= \calt \ket{\phi_1} \mcomma \\
\mathbb{E}_3 \ket{\psi_4} &= \balt \ket{\psi_3} \mcomma & \qquad  \mathbb{F}_2 \ket{\psi_4} &= d \ket{\phi_2} \mcomma \\
\mathbb{E}_2 \ket{\psi_3} &= b \ket{\phi_1} \mcomma & \qquad  \mathbb{F}_3 \ket{\psi_3} &= \dalt \ket{\psi_4} \mcomma
\end{aligned}
\end{equation}
where $\aalt,\balt,\calt,\dalt,a,b,c,d$ are coefficients constrained by the commutation relations of the $q$-deformed algebra. By renormalizing the states one could in principle eliminate two of the coefficients with hat, setting for instance $\aalt = \balt=1$, but here we prefer to keep the coefficients free, while ensuring the normalization $\bra{\phi_a}\ket{\phi_a}=\bra{\psi_\alpha}\ket{\psi_\alpha}=1$. Choosing the basis of states
\begin{equation}
\label{eq:su22basisvectors}
\ket{\phi_1}= \begin{pmatrix}
1 \\ 0 \\ 0 \\ 0
\end{pmatrix} \mcomma \qquad \ket{\phi_2}= \begin{pmatrix}
0 \\ 1 \\ 0 \\ 0
\end{pmatrix} \mcomma \qquad \ket{\psi_4}= \begin{pmatrix}
0 \\ 0 \\ 1 \\ 0
\end{pmatrix} \mcomma \qquad \ket{\psi_3}= \begin{pmatrix}
0 \\ 0 \\ 0 \\ 1
\end{pmatrix} \mcomma
\end{equation}
the positive and negative simple roots have matrix realizations
\begin{equation}
\label{eq:rootmatrixrealization} \begin{aligned}
\mathbb{E}_1 = \begin{pmatrix}
0 & 0 & 0 & 0\\
\aalt & 0 & 0 & 0\\
0 & 0 & 0 & 0\\
0 & 0 & 0 & 0
\end{pmatrix} \mcomma \qquad \mathbb{E}_2 = \begin{pmatrix}
0 & 0 & 0 & b\\
0 & 0 & 0 & 0\\
0 & a & 0 & 0\\
0 & 0 & 0 & 0
\end{pmatrix} \mcomma \qquad \mathbb{E}_3 = \begin{pmatrix}
0 & 0 & 0 & 0\\
0 & 0 & 0 & 0\\
0 & 0 & 0 & 0\\
0 & 0 & \balt & 0
\end{pmatrix} \mcomma \\
\mathbb{F}_1 = \begin{pmatrix}
0 & \calt & 0 & 0\\
0 & 0 & 0 & 0\\
0 & 0 & 0 & 0\\
0 & 0 & 0 & 0
\end{pmatrix} \mcomma \qquad \mathbb{F}_2 = \begin{pmatrix}
0 & 0 & 0 & 0\\
0 & 0 & d & 0\\
0 & 0 & 0 & 0\\
c & 0 & 0 & 0
\end{pmatrix} \mcomma \qquad \mathbb{F}_3 = \begin{pmatrix}
0 & 0 & 0 & 0\\
0 & 0 & 0 & 0\\
0 & 0 & 0 & \dalt\\
0 & 0 & 0 & 0
\end{pmatrix} \mdot
\end{aligned}
\end{equation}
The matrix realization of the other generators can easily be deduced from their expressions in terms of $\mathbb E_j$ and $\mathbb F_j$. Taking the commutator between a positive and a negative simple root one obtains the relations
\begin{gather}
\label{eq:closure1}
\aalt \calt = 1 \mcomma \qquad \balt \dalt = 1 \mcomma
\shortintertext{and}
\label{eq:closure2}
a d = \Q{C+\frac{1}{2}} , \qquad b c = \Q{C-\frac{1}{2}} .
\end{gather}
The commutation relations involving a Cartan element and a positive or negative simple root are automatically satisfied. Furthermore, the central charges $P$ and $K$, expectation values of the central elements $\mathbb P$ and $\mathbb K$ respectively, are given by
\begin{equation}
P = a b \aalt \balt \mcomma \qquad K = c  d \calt \dalt \mdot
\end{equation}
This in turn implies the closure condition
\begin{equation}
 \Q{C+\frac{1}{2}} \Q{C-\frac{1}{2}} = PK =  \beta^2 (1-q^{2 C} U^2)(q^{-2 C} -U^{-2}) \mcomma
\end{equation}
where for the last equality we plugged in the explicit expressions for $P$ and $K$ derived in \eqref{eq:PK_oxo}. This can be recast into
\begin{equation}
\label{eq:closure}
(V-V^{-1})^2 = \xi^2 (U-U^{-1})^2 + (1-\xi^2)(q^{1/2}-q^{-1/2})^2 \mcomma
\end{equation}
where we introduced
\begin{equation}
\label{eq:Vxi}
V= q^C \mcomma \qquad \xi = -i \frac{\beta (q-q^{-1})}{\sqrt{1-\beta^2 (q-q^{-1})^2}} \mdot
\end{equation}
The labeling of states by $\phi$ and $\psi$ as used in this section (and similarly in \cite{Beisert:2008tw}), corresponds in our conventions to the sigma model indices $1,2,3,4$ as
\begin{equation}
\label{eq:indexmatching}
\begin{alignedat}{2}
	\phi_1 & \leftrightarrow 1
	\mcomma \qquad &
	\psi_3 & \leftrightarrow 3
	\mcomma \\
	\phi_2 & \leftrightarrow 2
	\mcomma \qquad &
	\psi_4 & \leftrightarrow 4
	\mcomma
\end{alignedat}
\end{equation}
with a second copy of these for the dotted indices.

\paragraph{Deformation of the Zhukovsky variables.}
As customary in the context of $q$-de\-form\-a\-tions we introduce deformations of the Zhukovsky variables,
\begin{equation}
\label{eq:UV}
U^2 = q^{-1} \frac{x^+ + \xi}{x^- + \xi} =  q \frac{1/x^-+\xi}{1/x^+ + \xi} \mcomma \qquad
V^2 = q^{-1} \frac{1+ x^+ \xi}{1+ x^- \xi} = q \frac{\xi/x^- + 1}{\xi/x^+ +1} \mcomma
\end{equation}
and hence in the $x^\pm$ variables the closure condition \eqref{eq:closure} becomes
\begin{equation}
\label{eq:eqq}
q^{-1} \left( x^+ + \frac{1}{x^+} \right)- q \left( x^- + \frac{1}{x^-} \right) -(q-q^{-1}) \left(\xi + \frac{1}{\xi}\right)=0 \mdot
\end{equation}
The variables $\aalt$ and $\balt$ remain free, while the others are
\begin{equation}
\begin{aligned}
\calt&= \frac{1}{\aalt} \mcomma \qquad \qquad \qquad \qquad \dalt = \frac{1}{\balt} \mcomma \\
a &= \sqrt{\beta} \gamma U^{-1/2} q^{1/2}  \frac{1}{\aalt} \mcomma \\
b &= \sqrt{\beta} \alpha \gamma^{-1} U^{-1/2} \left(1-\frac{x^+}{x^-} \right) q^{-1/2} \frac{1}{\balt} \mcomma \\
c &= i\sqrt{\beta} \alpha^{-1} \gamma \sqrt{1-\xi^2} q U^{1/2} V^{-1} \frac{1}{x^++\xi} \balt \mcomma \\
d &= i\sqrt{\beta} \gamma^{-1} \sqrt{1-\xi^2} U^{1/2}  V \frac{x^- - x^+}{1+x^+ \xi} \aalt \mdot
\end{aligned}
\end{equation}
\paragraph{The $\mathfrak{sl}_q(2|2)_\text{c.e.}$ S matrix.} The S matrix satisfying the defining equality \eqref{eq:S_oxo} is \cite{Beisert:2008tw}%
\footnote{The S matrix acts in the tensor product of two spaces, which we label with indices $1,2$. (Not to be confused with the $\mathfrak{sl}(2)$ indices of the states $\ket{\phi_{1,2}}$.) Due to our choice of normalization of fields, the S matrix is obtained from \cite{Beisert:2008tw} by sending $\ket{\phi_2} \to q^{-1/2}\aalt \ket{\phi_2}$ and $\ket{\psi_1} \to q^{1/2}\balt \ket{\psi_3}$.}
\begin{equation*}
\label{eq:exactSdistinguished}
\resizeToFitPageText{
	\everydisplay={
		\settowidth{\displaywidth}{$
			\displaystyle
			S \ket{\psi_3 \psi_4} = -\frac{\balt_2}{\balt_1}\frac{q^2 \SD + \SE}{q^2+1} \ket{\psi_4 \psi_3} - \frac{q(\SD - \SE)}{q^2+1} \ket{\psi_3 \psi_4} + \frac{\aalt_1}{\balt_1} \frac{ q^{-1} \SF}{q^2+1} \ket{\phi_2 \phi_1} - \frac{\aalt_2}{\balt_1} \frac{\SF}{q^2+1} \ket{\phi_1 \phi_2} ,
		$}
	}
	\begin{align*}
		&S \ket{\phi_a \phi_a} = \SA \ket{\phi_a \phi_a} , \qquad \qquad \qquad \qquad \qquad \quad S \ket{\psi_\alpha \psi_\alpha} = -\SD \ket{\psi_\alpha \psi_\alpha} ,
		\\[7pt]
		&S \ket{\phi_1 \phi_2} = \frac{\aalt_1}{\aalt_2}\frac{q^2 \SA + \SB}{q^2+1} \ket{\phi_2 \phi_1} + \frac{q(\SA - \SB)}{q^2+1} \ket{\phi_1 \phi_2} - \frac{\balt_2}{\aalt_2} \frac{q \SC}{q^2+1} \ket{\psi_4 \psi_3} + \frac{\balt_1}{\aalt_2} \frac{q^ 2 \SC}{q^2+1} \ket{\psi_3 \psi_4} , \\
		&S \ket{\phi_2 \phi_1} = \frac{q(\SA - \SB)}{q^2+1} \ket{\phi_2 \phi_1} + \frac{\aalt_2}{\aalt_1} \frac{ \SA + q^2 \SB}{q^2+1} \ket{\phi_1 \phi_2} + \frac{\balt_2}{\aalt_1} \frac{q^2 \SC}{q^2+1} \ket{\psi_4 \psi_3} - \frac{\balt_1}{\aalt_1} \frac{q^3 \SC}{q^2+1} \ket{\psi_3 \psi_4} , \\
		&S \ket{\psi_3 \psi_4} = -\frac{\balt_2}{\balt_1}\frac{q^2 \SD + \SE}{q^2+1} \ket{\psi_4 \psi_3} - \frac{q(\SD - \SE)}{q^2+1} \ket{\psi_3 \psi_4} + \frac{\aalt_1}{\balt_1} \frac{ q^{-1} \SF}{q^2+1} \ket{\phi_2 \phi_1} - \frac{\aalt_2}{\balt_1} \frac{\SF}{q^2+1} \ket{\phi_1 \phi_2} , \\
		&S \ket{\psi_4 \psi_3} = -\frac{q(\SD - \SE)}{q^2+1} \ket{\psi_4 \psi_3} - \frac{\balt_1}{\balt_2} \frac{\SD + q^2 \SE}{q^2+1} \ket{\psi_3 \psi_4} - \frac{\aalt_1}{\balt_2} \frac{ \SF}{q^2+1} \ket{\phi_2 \phi_1} + \frac{\aalt_2}{\balt_2} \frac{q \SF}{q^2+1} \ket{\phi_1 \phi_2} ,
		\\[3pt]
		&\begin{aligned}[t]
			S \ket{\phi_1 \psi_3} &= \SG \ket{\phi_1 \psi_3} + \frac{\balt_1}{\balt_2} \SH \ket{\psi_3 \phi_1} , &\qquad \quad S \ket{\psi_3 \phi_1} &= \mathrlap{\SL \ket{\psi_3 \phi_1} + \frac{\balt_2}{\balt_1} \SK  \ket{\phi_1 \psi_3} ,} \\
			S \ket{\phi_1 \psi_4} &= \SG \ket{\phi_1 \psi_4} + \SH \ket{\psi_4 \phi_1} , &\qquad \quad S \ket{\psi_4 \phi_1} &= \mathrlap{\SL \ket{\psi_4 \phi_1} + \SK  \ket{\phi_1 \psi_4} ,} \\
			S \ket{\phi_2 \psi_3} &= \SG \ket{\phi_2 \psi_3} + \frac{\aalt_2 \balt_1}{\aalt_1 \balt_2} \SH \ket{\psi_3 \phi_2} , &\qquad \quad S \ket{\psi_3 \phi_2} &= \mathrlap{\SL \ket{\psi_3 \phi_2} + \frac{\balt_2 \aalt_1}{\balt_1 \aalt_2}  \SK \ket{\phi_2 \psi_3} ,} \\
			S \ket{\phi_2 \psi_4} &= \SG \ket{\phi_2 \psi_4} + \frac{\aalt_2}{\aalt_1} \SH \ket{\psi_4 \phi_2} ,&\qquad \quad S \ket{\psi_4 \phi_2} &= \mathrlap{\SL \ket{\psi_4 \phi_2} + \frac{\aalt_1}{\aalt_2}  \SK \ket{\phi_2 \psi_4} .}
		\end{aligned}
		\numberthis 
	\end{align*}
}
\end{equation*}
The ten coefficients are given by
\begin{equation}
\begin{aligned}
\SA &= S_0 \frac{U_1 V_1}{U_2 V_2} \frac{x_2^+ - x_1^-}{x_2^- - x_1^+} \mcomma \\
\SB &= S_0 \frac{U_1 V_1}{U_2 V_2} \frac{x_2^+ - x_1^-}{x_2^- - x_1^+} \left( 1-(q+q^{-1})q^{-1} \frac{x_2^+ - x_1^+}{x_2^+ - x_1^-} \frac{x_2^- - 1/x_1^+}{x_2^- - 1/x_1^-}\right) \mcomma\\
\SC &= -S_0 (q+q^{-1}) \frac{ \gamma_1 \gamma_2 U_1 V_1}{\alpha q^{3/2} U_2^2 V_2^2}\frac{x_1^-}{x_1^+} \frac{x_1^+ - x_2^+}{(x_1^+ - x_2^-)(1- x_1^- x_2^-)} \mcomma \\
\SD &= -S_0 \mcomma \\
\SE &= -S_0  \left( 1-(q+q^{-1}) \frac{1}{q U_2^2 V_2^2} \frac{x_2^+ - x_1^+}{x_2^- - x_1^+} \frac{x_2^+ - 1/x_1^-}{x_2^- - 1/x_1^-}\right) \mcomma \\
\SF &= - S_0 (q+q^{-1}) \frac{\alpha U_1^2 V_1^2}{q^{1/2} U_2 V_2 \gamma_1 \gamma_2} \frac{x_1^-}{x_1^+} \frac{(x_1^- - x_1^+)(x_2^+ - x_1^+)(x_2^+ - x_2^-)}{(x_2^- - x_1 ^+)(1-x_1^- x_2^-)} \mcomma \\
\SG &= S_0 \frac{1}{q^{1/2} U_2 V_2} \frac{x_2^+ - x_1^+}{x_2^- - x_1^+} \mcomma \\
\SH &= S_0  \frac{\gamma_1}{\gamma_2} \frac{x_2^+ - x_2^-}{x_2^- - x_1^+} \mcomma \\
\SK &= S_0 \frac{U_1 V_1}{U_2 V_2} \frac{\gamma_2}{\gamma_1} \frac{x_1^+ - x_1^-}{x_2^- - x_1^+} \mcomma \\
\SL &= S_0 U_1 V_1 q^{1/2} \frac{x_2^- - x_1^-}{x_2^- - x_1 ^+} \mdot
\end{aligned}
\end{equation}
This S matrix satisfies the quantum Yang-Baxter equation%
\footnote{
	The $S_{ij}$ denote the graded embeddings of $S$,
	defined analogously to the $\dsT_{ij}$ of \cref{definition-graded-embeddings-T}.
}
\begin{equation}
	S_{12} S_{13} S_{23} = S_{23} S_{13} S_{12} \mdot
\end{equation}

\subsubsection{Distinguished Dynkin diagram}
Let us briefly discuss the exact $\mathfrak{su}_q(2|2)_\mup{c.e.}$ S matrix associated with the distinguished Dynkin diagram.
\paragraph{Reality conditions.} In order to obtain the exact S matrix for the deformed model based on the distinguished Dynkin diagram we need to impose the reality conditions \eqref{eq:rc_oxo}. This implies $\xi \in i \, \mathbb{R}$ together with the constraints
\begin{equation} \begin{aligned}
&\abs{\aalt}^2 = q \mcomma &\qquad &\abs*{\balt}^2 = q^{-1} \mcomma \\
&\abs{\gamma}^2 = i (1-q V^2) \frac{\smash{\sqrt{1-\xi^2}}}{\xi} \mcomma& \qquad  &\abs{\alpha}^2=1 \mdot \end{aligned}
\end{equation}
A solution to these equations is given by%
\footnote{
\label{explanation-phase-gamma}%
In \cite{Beisert:2008tw}, $\gamma$ was instead taken to be $\gamma = \sqrt{- i q^{1/2} U V (x^+-x^-)}$, which differs by an overall factor from our choice \eqref{eq:rc_oxo_val}. This factor can be reabsorbed into a renormalization (by a phase) of the fermionic states $\ket{\psi_3}$ and $\ket{\psi_4}$ to yield the same S matrix. The reason we choose this particular value of $\gamma$ is to exactly reproduce the perturbative T matrix upon expansion without having to renormalize the fermions, cf.\ \cref{explanation-phase-f}.
}
\begin{equation}
\label{eq:rc_oxo_val}
\aalt = q^{1/2} \mcomma \qquad \balt = q^{-1/2} \mcomma \qquad \alpha =1 \mcomma \qquad  \gamma = \sqrt{-q^{1/2} U V (x^+-x^-)} \mdot
\end{equation}

\paragraph{Crossing symmetry.} A virtue of the coefficients \eqref{eq:rc_oxo_val} is that they lead to an exact S matrix which has crossing symmetry
\newcommand{\supertranspose}{{\scriptscriptstyle \textsf{ST}}}
\begin{equation}
\label{eq:crossing}
(\mathcal C^{-1} \otimes 1) S_{\bar 1 2}^{\supertranspose \otimes 1} (\mathcal C \otimes 1) S_{12} =1 \otimes 1 \mcomma
\end{equation}
provided the overall prefactor satisfies the crossing relation
\begin{equation}
\label{eq:crossingS0}
(S_0)_{12} (S_0)_{\bar 1 2} = q \frac{x_2^+}{x_2^-} \frac{(x_1^- - x_2^ -)(1-x_1^+ x_2^-)}{(x_1^--x_2^+)(1-x_1^+ x_2^+)} \mdot
\end{equation}
In \eqref{eq:crossing}, $-^{\supertranspose}$ denotes the supertransposition acting as $A^{\supertranspose}_{jk} = (-1)^{(\abs{j}+1) \abs{k}} A_{kj}$, with $\abs{j}=0$ if the index is bosonic and $\abs{j}=1$ if the index is fermionic, $\bar 1$ means that we consider the antipode representation,
\begin{equation}
\label{eq:antipoderep}
\bar x^\pm = \frac{1}{x^\pm} \mcomma \qquad \bar U = U^{-1} \mcomma \qquad \bar V = V^{-1} \mcomma \qquad  \gamma \bar \gamma = -i q^{1/2}  (UV-U^{-1}V^{-1}) \mcomma
\end{equation}
and the charge conjugation matrix is given by%
\footnote{
\label{footnote:chargeconjrelation}%
The charge conjugation matrix is defined through the relation
$\antipode(X) = \chargeconj^{-1} \bar{X}^ {\supertranspose} \chargeconj \mcomma$
with $\antipode$ the antipode map defined in \cref{eq:antipode_oxo}. Imposing this relation for the bosonic simple roots $\mathbb E_1, \mathbb F_1, \mathbb E_3$ and $\mathbb F_3$ fixes the conjugation matrix up to an overall prefactor in the two subspaces $\{\ket{\phi_1},\ket{\phi_2}\}$ and $\{\ket{\psi_3},\ket{\psi_4}\}$. Further requiring that it holds for the fermionic simple roots $\mathbb E_2$ and $\mathbb F_2$ fixes one prefactor in terms of the other and one obtains the antipode representation for the parameter $x^\pm$ and $\gamma$ as in \cref{eq:antipoderep}. Notice that since our $\gamma$ differs from the one chosen in \cite{Beisert:2008tw}, so does the charge conjugation matrix.
}
\begin{equation}
\label{eq:chargeconj_oxo}
\begin{aligned}
\chargeconj \ket{\phi_1} &= - q^{1/2} \ket{\phi_2} \mcomma &\qquad \chargeconj \ket{\psi_3} &= -i q^{1/2} \ket{\psi_4} \mcomma \\
\chargeconj \ket{\phi_2} &= + q^{-1/2} \ket{\phi_1} \mcomma &\qquad \chargeconj \ket{\psi_4} &= + i q^{-1/2} \ket{\psi_3} \mdot
\end{aligned}
\end{equation}

\paragraph{A symmetry of the distinguished exact S matrix.} An interesting property of the exact S matrix for the distinguished Dynkin diagram with \eqref{eq:rc_oxo_val} is its invariance under the map
\begin{equation}
\label{eq:sym_exact_dist}
q \rightarrow  q^{-1} , \qquad \ket{\phi_1} \leftrightarrow \ket{\phi_2} , \qquad \ket{\psi_3} \leftrightarrow \ket{\psi_4} .
\end{equation}
To show this, let us analyze the consequences of sending $q \rightarrow q^{-1}$ on the coefficients of the S matrix. First of all, the braiding factor $U$ is independent on $q$ and thus remains unchanged. By the definitions \eqref{eq:Vxi} we have $V \rightarrow V^{-1}$ and $\xi \rightarrow -\xi$. The variables $x^\pm$ also need to be modified as they are subject to the conditions \eqref{eq:UV} and \eqref{eq:eqq}, which depend on $q$, $V$ and $\xi$. The solutions to the modified constraints are
\begin{equation}
x^- \rightarrow \frac{x^- + \xi}{1+x^- \xi} \mcomma \qquad x^+ \rightarrow \frac{x^+ + \xi}{1+x^+ \xi} \mdot
\end{equation}
This in turn implies
\begin{equation}
\gamma \rightarrow \gamma \frac{\sqrt{1-\xi^2}}{1+x^+ \xi} \mdot
\end{equation}
Under these transformations the right-hand side of \eqref{eq:crossingS0} remains invariant. Moreover, also the ten coefficients $\SA,\SB, \dots, \SL$ are left unchanged. The latter however enter the S matrix with factors of the deformation parameter $q$. Therefore the transformation $q \rightarrow q^{-1}$ is not itself a symmetry of the exact S matrix associated to the distinguished Dynkin diagram, but it is easy to see that it becomes one when supplemented with the exchange of states as in \eqref{eq:sym_exact_dist}.

\subsubsection{Fermionic Dynkin diagram}
\label{exact-fermionic-S-matrix}
We now consider the $\mathfrak{su}_q(2|2)_\mup{c.e.}$ S matrix associated to the fermionic Dynkin diagram.

\paragraph{Implementing the twist.} In order to obtain the exact S matrix $S'$ associated to the fermionic Dynkin diagram we implement the twist \eqref{eq:Smatrix_xxx}, $S'=\Ftwist^{-\mup{op}} S \Ftwist$. In the fundamental representation, $F$ only differs from the identity for the following matrix elements:
\begin{equation} \begin{aligned}
(\Ftwist-1\otimes 1) \ket{\phi_1 \phi_2} &=  -(q-q^{-1}) U_1^{1/2} U_2^{1/2} c_1 a_2 \ket{\psi_3 \psi_4} \mcomma \\
(\Ftwist-1\otimes 1) \ket{\phi_1 \psi_3} &=  -(q-q^{-1}) U_1^{1/2} U_2^{1/2} c_1 b_2 \ket{\psi_3 \phi_1} \mcomma \\
(\Ftwist-1\otimes 1) \ket{\psi_4 \phi_2} &=  +(q-q^{-1}) U_1^{1/2} U_2^{1/2} d_1 a_2 \ket{\phi_2 \psi_4} \mcomma \\
(\Ftwist-1\otimes 1) \ket{\psi_4 \psi_3} &=  +(q-q^{-1}) U_1^{1/2} U_2^{1/2} d_1 b_2 \ket{\phi_2 \phi_1} \mdot\\
\end{aligned}
\end{equation}

\paragraph{Reality conditions.} Finally, to obtain a unitary S matrix, we impose the reality conditions \eqref{eq:rc_oxo_bis}, leading to
\begin{equation} \begin{aligned}
&\abs{\aalt}^2 = V^2 q^{-1} \mcomma &\qquad &\abs{\balt}^2 = V^2 q \mcomma \\
&\abs{\gamma}^2 =  i q^{-3} (1-q V^2) \frac{\sqrt{1-\xi^2}}{\xi} \mcomma& \qquad  &\abs{\alpha}^2=1 \mdot \end{aligned}
\end{equation}
With this choice of coefficients, the S matrix is unitary: $(S'_{12})^\dagger S'_{12} = 1 \otimes 1$, provided the overall prefactor is a pure phase, $\abs{S_0}=1$. A solution is given by
\begin{equation}
\label{eq:rc_xxx_val}
\aalt = V q^{-1/2} \mcomma \qquad \balt = V q^{1/2} \mcomma \qquad \alpha =1 \mcomma \qquad  \gamma = q^{-3/2} \sqrt{- q^{1/2} U V (x^+-x^-)} \mdot
\end{equation}

\paragraph{Crossing symmetry.}
The exact S matrix associated with the fermionic Dynkin diagram satisfies the crossing equation \eqref{eq:crossing}, with conjugation matrix%
\footnote{The charge conjugation is defined analogously to the relation of \cref{footnote:chargeconjrelation}. Since the twist does not preserve the antipode map, see \cref{eq:antipodetwist}, the charge conjugation here differs from the one in \eqref{eq:chargeconj_oxo}.}
\begin{equation}
\begin{aligned}
\chargeconj'\ket{\phi_1} &= - \ket{\phi_2} \mcomma &\qquad \chargeconj' \ket{\psi_3} &= -i \ket{\psi_4} \mcomma \\
\chargeconj' \ket{\phi_2} &= +\ket{\phi_1} \mcomma &\qquad \chargeconj' \ket{\psi_4} &= + i \ket{\psi_3} \mcomma
\end{aligned}
\end{equation}
and exactly the same relation \eqref{eq:crossingS0}. Therefore, if crossing symmetry is imposed, the prefactor is constrained in the same way for the distinguished and fermionic exact S matrix. It is thus consistent to pick it to be the same for both cases.

\subsubsection{Expansion of the exact S matrix}
\label{subsec:expansion_exact}

In order to compare to the perturbative result, we need to obtain the tree-level expansion of the fermionic $\mathfrak{su}_q(2|2)_\mup{c.e.}$ S matrix and provide a physical meaning to the purely algebraic quantities used to construct the exact fermionic S matrix. We assume that the energy and the momentum are defined as in the undeformed case through
\begin{equation}
\mathbb{C} \ket{\Phi} = C \ket{\Phi} =\frac{\omega}{2} \ket{\Phi} \mcomma \qquad \mathbb{U} \ket{\Phi} = U \ket{\Phi} = e^{\frac{i p}{2}} \ket{\Phi} \mcomma
\end{equation}
with $\Phi$ standing for an element of $\{ \phi_1,\phi_2,\psi_3,\psi_4  \}$. The exact S matrix has two free parameters $q$ and $\xi$. Based on experience with the distinguished case \cite{Arutyunov:2013ega}, we take these to be related to the deformation parameter $\kappa$ and the string tension $\stringtension$ entering the string sigma model through\footnote{The expression for $\xi$ is equivalent to taking $\beta = \stringtension/(2\sqrt{1+\kappa^2})$. Taking $\stringtension = g \sqrt{1+\kappa^2}$, $\beta=g/2$ gives the expressions of \cite{Arutyunov:2013ega}.}
\begin{equation}
q = e^{-\kappa/\stringtension} \mcomma \qquad \xi = i \kappa \mdot
\end{equation}
Rescaling the momentum $p \rightarrow p/\stringtension$ we find to linear order in $\stringtension$
\begin{equation}
U = 1+ \frac{i p}{2 \stringtension}+\dots \mcomma \qquad V = 1 - \frac{\kappa \omega}{2 \stringtension}+\dots \mdot
\end{equation}
Solving the variables $x^\pm$ as functions of $U,V$, and using the closure condition \eqref{eq:closure} yields the dispersion relation \eqref{eq:dispersion}.
Finally, recalling that it is consistent to take the scalar factor $S_0$ to be equal to the one of the distinguished case \cite{Hoare:2011wr,Arutyunov:2013ega,Arutynov:2014ota} and matching indices as in eqs.\ \eqref{eq:indexmatching}, the expansion of the exact fermionic S matrix gives
\begin{equation}
\label{eq:exactSfactorexpansion}
S'(q) = \mathds{1} + \frac{i}{\stringtension} \mathcal{T}(\kappa) +\ldots \mcomma
\end{equation}
where we recover the matrix $\scT(\kappa)$ of \cref{definition-scT} used to express the perturbative result.

\section{Comparison of the perturbative and exact S matrix}
\label{sec:comparison}

In section \ref{T-matrix-factorization} we found a tree-level $\dsT$ matrix of the form
\begin{equation}
	\dsT = \scT(-\kappa) \otimes \identity + \identity \otimes \scT(\kappa)
	\mdot
\end{equation}
This structure is a deformation of the one for the undeformed string, which has two identical $\scT$ factors in its $\dsT$ matrix. It suggests that the two factors of the $\mathfrak{su}(2|2)_\mup{c.e.}^{\oplus2}$ light-cone symmetry of the undeformed string get deformed oppositely rather than identically. A closer look at the embedding of the relevant $\mathfrak{su}(2|2)$ algebras in $\mathfrak{su}(2,2|4)$, and the action of the $R$ operator at both levels, shows that this should indeed be the case.

\subsection{Light-cone symmetry algebra}

In the matrix conventions of \cite{Arutyunov:2009ga} (see e.g.\ equation (2.123) there), the two copies of $\mathfrak{su}(2|2)$ are embedded in $\mathfrak{su}(2,2|4)$ as
\begin{equation}
\label{eq:su22sinpsu224}
\left(
\begin{array}{cccc}
\mathbb{R} & 0 & -\mathbb{Q}^\dagger & 0 \\
0 & \mathring{\mathbb{R}} & 0 & \mathring{\mathbb{Q}} \\
\mathbb{Q} & 0 & \mathbb{L} & 0 \\
0 &  \mathring{\mathbb{Q}}^\dagger &0 & \mathring{\mathbb{L}}
\end{array}
\right),
\end{equation}
in $2\times2$ block notation, with one copy of $\mathfrak{su}(2|2)$ generated by $\mathbb{R}, \mathbb{L}, \mathbb{Q}, \mathbb{Q}^\dagger$, and the other by their dotted counterparts. Note the different relative placement of $\mathbb{Q}$ and $\mathring{\mathbb{Q}}$.\footnote{While $\mathbb{Q} \leftrightarrow \mathbb{Q}^\dagger$ is an automorphism of $\mathfrak{su}(2|2)$, the central extensions $\{\mathbb{Q},\mathbb{Q}\} \sim\mathbb{C}$ and $\{\mathring{\mathbb{Q}},\mathring{\mathbb{Q}}\}\sim\mathbb{C}$ that appear off shell, meaningfully fix this embedding.}

This structure needs to be contrasted with the action of the fermionic $R$ operator defining the action, and the fermionic $R$ operator $R_{\mathfrak{su}(2|2)}$ corresponding to the $q$ deformation of the exact  $\mathfrak{su}(2|2)$ S matrix. The $R$ operator defining the action, acts on elements $M$ of $\mathfrak{su}(2,2|4)$ as \cite{Hoare:2018ngg}
\begin{equation}
\label{eq:fermRaction}
R(M)_{ij} = - i \epsilon_{ij} M_{ij} \mcomma \qquad \epsilon =
\left(
\begin{array}{cc;{2pt/4pt}cc;{2pt/4pt}cc;{2pt/4pt}cc}
\cellcolor{color1} 0 & \cellcolor{color1}  +1 & +1 &+1 & \cellcolor{color1}+1 &  \cellcolor{color1}+1 & +1 & +1 \\
\cellcolor{color1} -1 & \cellcolor{color1} 0 & +1 & +1 &  \cellcolor{color1}-1 &  \cellcolor{color1}+1 & +1 & +1 \\
\hdashline[2pt/4pt]
-1 & -1 & \cellcolor{color2} 0 & \cellcolor{color2}+1 & -1 & -1 &  \cellcolor{color2}+1 &  \cellcolor{color2}+1 \\
-1 & -1 &\cellcolor{color2} -1 & \cellcolor{color2}0 & -1 & -1 &  \cellcolor{color2}-1 &  \cellcolor{color2}+1 \\
\hdashline[2pt/4pt]
\cellcolor{color1} -1 &  \cellcolor{color1}+1 & +1 & +1 & \cellcolor{color1} 0 & \cellcolor{color1} +1 & +1 & +1 \\
\cellcolor{color1} -1 &  \cellcolor{color1}-1 & +1 & +1 & \cellcolor{color1} -1 & \cellcolor{color1} 0 & +1 & +1 \\ \hdashline[2pt/4pt]
-1 & -1 &  \cellcolor{color2}-1 &  \cellcolor{color2}+1 & -1 & -1 & \cellcolor{color2} 0 & \cellcolor{color2}+1 \\
-1 & -1 &  \cellcolor{color2}-1 &  \cellcolor{color2}-1 & -1 & -1 & \cellcolor{color2}-1 & \cellcolor{color2}0
\end{array}
\right),
\end{equation}
where we have highlighted the blocks corresponding to the undotted and dotted copies of $\mathfrak{su}(2|2)$ in green and yellow respectively. In appendix \ref{app:su22Roperators} we translate between conventions to determine the $R$ operator corresponding to the exact S matrix of section \ref{sec:exactSmatrix}, acting on a copy of $\mathfrak{su}(2|2)$ of the form
\begin{equation}
\left(
\begin{array}{cc}
\mathbb{R} & -\mathbb{Q}^\dagger \\
\mathbb{Q} & \mathbb{L}
\end{array}
\right).
\end{equation}
This $R$ operator is
\begin{equation}
R_{\mathfrak{su}(2|2)}(M)_{ij} = - i \epsilon_{ij} M_{ij} \mcomma \qquad \epsilon =
\left(
\begin{array}{cc;{2pt/4pt}cc}
0 & -1 & -1 & -1 \\
+1 & 0 & +1 & -1\\
\hdashline[2pt/4pt]
+1 & -1 & 0 & -1\\
+1 & +1 & +1 & 0
\end{array}
\right).
\end{equation}

Comparing the action of $R_{\mathfrak{su}(2|2)}$ to the action of $R$ on the two $\mathfrak{su}(2|2)$s as in equation \eqref{eq:fermRaction}, we see that $R$ acts like $-R_{\mathfrak{su}(2|2)}$ on the undotted copy of $\mathfrak{su}(2|2)$. For the dotted copy we can now compare term by term at the level of the individual indexed generators (see e.g.\ section 2.4.2 of \cite{Arutyunov:2009ga}).\footnote{Recall that we permute indices $1$ and $2$, and $\dot{3}$ and $\dot{4}$, relative to \cite{Arutyunov:2009ga}.} In terms of index assignment, \eqref{eq:su22sinpsu224} has the form
\begin{equation}
\label{eq:RactionvsIndicesFermionic}
\addtolength{\arraycolsep}{-1pt}
\left(
\begin{array}{cccccccc}
34&  {33}\cellcolor{color3} & \cellcolor{color3} & \cellcolor{color3} & 24\cellcolor{color3} & 14\cellcolor{color3} & \cellcolor{color3} & \cellcolor{color3} \\
44 \cellcolor{color4} &  43 & \cellcolor{color3} & \cellcolor{color3} & 23\cellcolor{color4} & 13\cellcolor{color3} & \cellcolor{color3} & \cellcolor{color3} \\
\cellcolor{color4} & \cellcolor{color4} & \dot{4}\dot{3} & \dot{4}\dot{4}\cellcolor{color3} & \cellcolor{color4} & \cellcolor{color4} & \dot{4}\dot{2}\cellcolor{color3} & \dot{4}\dot{1}\cellcolor{color3} \\
\cellcolor{color4} & \cellcolor{color4} & \dot{3}\dot{3}\cellcolor{color4} & \dot{3}\dot{4}  & \cellcolor{color4} & \cellcolor{color4} & \dot{3}\dot{2}\cellcolor{color4} & \dot{3}\dot{1}\cellcolor{color3} \\
42\cellcolor{color4} & 32\cellcolor{color3} & \cellcolor{color3} & \cellcolor{color3} & 21  &  22\cellcolor{color3} & \cellcolor{color3} & \cellcolor{color3} \\
41\cellcolor{color4} & 31\cellcolor{color4} & \cellcolor{color3} & \cellcolor{color3} &  11\cellcolor{color4} & 12 & \cellcolor{color3} & \cellcolor{color3} \\
\cellcolor{color4} & \cellcolor{color4} & \dot{2}\dot{4}\cellcolor{color4} &\dot{2}\dot{3} \cellcolor{color3} & \cellcolor{color4} & \cellcolor{color4} &\dot{1}\dot{2}  & \dot{1}\dot{1}\cellcolor{color3} \\
\cellcolor{color4} & \cellcolor{color4} & \dot{1}\dot{4}\cellcolor{color4} &\dot{1}\dot{3} \cellcolor{color4} & \cellcolor{color4} & \cellcolor{color4} & \dot{2}\dot{2}\cellcolor{color4} & \dot{2}\dot{1}
\end{array}
\right)
\end{equation}
where we have indicated the action of the $R$ operator by color-coding the entries in red $(+i)$, blue $(-i)$ and white $(0)$. We see that $R$ acts precisely oppositely on the dotted and undotted (indexed) generators of the two copies of $\mathfrak{su}(2|2)$. Since $R$ acted like $-R_{\mathfrak{su}(2|2)}$ on the undotted $\mathfrak{su}(2|2)$, it acts like $+R_{\mathfrak{su}(2|2)}$ on the dotted $\mathfrak{su}(2|2)$ with permuted indices.

As changing the sign of the $R$ operator is equivalent to changing the sign of $\kappa$ or inverting $q$, the two copies of $\mathfrak{su}(2|2)$ effectively have opposite deformation parameters. Putting everything together, our light-cone symmetry algebra is thus expected to be $\mathfrak{su}_{1/q}(2|2)_\mup{c.e.} \oplus \mathfrak{su}_{q}(2|2)_\mup{c.e.}$.

\subsection{Expanded exact S matrix}

The above structure for the light-cone symmetry algebra is compatible with our tree-level $\dsT$ matrix, and suggest that the exact S matrix is of the form $S^\prime(1/q)\otimes {S^\prime}(q)$, where the factors correspond to the fermionic exact $\mathfrak{su}_{q}(2|2)_\mup{c.e.}$ S matrix $S^\prime(q)$. Using the tree-level expansion of $S^\prime(q)$ given in equation \eqref{eq:exactSfactorexpansion}, we find
\begin{equation}
S^\prime(1/q)\otimes {S^\prime}(q) = \mathds{1} + \frac{i}{\stringtension} \left( \scT(-\kappa) \otimes \identity + \identity \otimes \scT(\kappa)\right) + \ldots = \mathds{1} + \frac{i}{\stringtension} \dsT +\ldots \mdot
\end{equation}
In other words we find perfect agreement between the exact $\mathfrak{su}_{1/q}(2|2)_\mup{c.e.} \oplus \mathfrak{su}_{q}(2|2)_\mup{c.e.}$ S matrix and our tree-level $\dsT$ matrix, provided $q=e^{-\kappa/\stringtension}$, at least semiclassically. Taking into account the relative parametrizations, this identification of $q$ matches the one of \cite{Delduc:2014kha}.

\section{Distinguished deformation}
\label{sec:distinguishedcase}

With the framework set up, we can quickly revisit the computation of the tree-level $\dsT$ matrix for the distinguished deformation.
Taking the distinguished background presented in \cite{Arutyunov:2015qva} as input, we proceed exactly as described before, except that we use the unconjugated $f_p$ and $h_p$ in the mode expansion of $\theta$
\begin{fleqn}
\begin{equation}
	\theta^{a \dot{\alpha}}(\tau, \sigma) =
	\frac{\E^{-\I \pi/4}}{\sqrt{2\pi}} \int \dd{p} \frac{1}{\sqrt{\omega_p}}
	\qty(
		- \I \E^{\I (p \sigma - \omega_p \tau)} f_p a^{a \dot{\alpha}}(p)
		- \I \E^{-\I (p \sigma - \omega_p \tau)} h_p \epsilon^{a b} \epsilon^{\dot{\alpha} \dot{\beta}} a^\dagger_{b \dot{\beta}}(p)
	)
	\mathrlap{,}
\end{equation}
\end{fleqn}
for convenient direct comparison. The resulting $\dsT$ matrix factorizes, and solves the CYBE. As for the fermionic case, it is of the form
\begin{equation}
	\dsT = \scT(-\kappa) \otimes \identity + \identity \otimes \scT(\kappa)
\end{equation}
where $\scT$ is now given by \cref{definition-scT} with $\Aalgebraic$, $\Balgebraic$, $\Galgebraic$ and $\Walgebraic$ as in the fermionic case, see \cref{T-matrix-coefficients}, and
\begin{equation}
	\Calgebraic_{a b}^{\alpha \beta}(\kappa) = \CbarAlgebraic_{\alpha \beta}^{a b}(\kappa) = \Ccommon
	\mcomma \qquad
	\Halgebraic_{a \beta}^{\alpha b}(\kappa) = \HbarAlgebraic_{\alpha b}^{a \beta}(\kappa) = \Hcommon
	\mdot
\end{equation}

Although the $R$ operator is different, the light-cone symmetry algebra has the same overall structure as in the fermionic case. As discussed in appendix \ref{app:su22Roperators}, we now have
\begin{equation}
R_{\mathfrak{su}(2|2)}(M)_{ij} = - i \epsilon_{ij} M_{ij} \mcomma \qquad \epsilon =
\left(
\begin{array}{cc;{2pt/4pt}cc}
0 & -1 & -1 & -1 \\
+1 & 0 & -1 & -1\\
\hdashline[2pt/4pt]
+1 & +1 & 0 & -1\\
+1 & +1 & +1 & 0
\end{array}
\right),
\end{equation}
and the analogue of \eqref{eq:RactionvsIndicesFermionic} becomes
\begin{equation}
\label{eq:RactionvsIndicesDistinguished}
\addtolength{\arraycolsep}{-1pt}
\left(
\begin{array}{cccccccc}
34&  {33}\cellcolor{color3} & \cellcolor{color3} & \cellcolor{color3} & 24\cellcolor{color3} & 14\cellcolor{color3} & \cellcolor{color3} & \cellcolor{color3} \\
44 \cellcolor{color4} &  43 & \cellcolor{color3} & \cellcolor{color3} & 23\cellcolor{color3} & 13\cellcolor{color3} & \cellcolor{color3} & \cellcolor{color3} \\
\cellcolor{color4} & \cellcolor{color4} & \dot{4}\dot{3} & \dot{4}\dot{4}\cellcolor{color3} & \cellcolor{color3} & \cellcolor{color3} & \dot{4}\dot{2}\cellcolor{color3} & \dot{4}\dot{1}\cellcolor{color3} \\
\cellcolor{color4} & \cellcolor{color4} & \dot{3}\dot{3}\cellcolor{color4} & \dot{3}\dot{4}  & \cellcolor{color3} & \cellcolor{color3} & \dot{3}\dot{2}\cellcolor{color3} & \dot{3}\dot{1}\cellcolor{color3} \\
42\cellcolor{color4} & 32\cellcolor{color4} & \cellcolor{color4} & \cellcolor{color4} & 21  &  22\cellcolor{color3} & \cellcolor{color3} & \cellcolor{color3} \\
41\cellcolor{color4} & 31\cellcolor{color4} & \cellcolor{color4} & \cellcolor{color4} &  11\cellcolor{color4} & 12 & \cellcolor{color3} & \cellcolor{color3} \\
\cellcolor{color4} & \cellcolor{color4} & \dot{2}\dot{4}\cellcolor{color4} &\dot{2}\dot{3} \cellcolor{color4} & \cellcolor{color4} & \cellcolor{color4} &\dot{1}\dot{2}  & \dot{1}\dot{1}\cellcolor{color3} \\
\cellcolor{color4} & \cellcolor{color4} & \dot{1}\dot{4}\cellcolor{color4} &\dot{1}\dot{3} \cellcolor{color4} & \cellcolor{color4} & \cellcolor{color4} & \dot{2}\dot{2}\cellcolor{color4} & \dot{2}\dot{1}
\end{array}
\right)\,.
\end{equation}
We see that also in the distinguished case we expect $\mathfrak{su}_{1/q}(2|2)_\mup{c.e.} \oplus \mathfrak{su}_{q}(2|2)_\mup{c.e.}$ symmetry. Expanding the corresponding exact S matrix reproduces our tree-level result here as well.

The inversion of the deformation parameter is less significant here than it was in the fermionic case. As discussed around equations \eqref{eq:sym_exact_dist}, for the distinguished deformation an inversion of the deformation parameter is equivalent to a change of basis
\begin{equation}
	S(1/q)=\permute{S}(q) \quad \implies  \quad \scT(-\kappa)=\permute{\scT}(\kappa),
\end{equation}
where $\permute{-}$ denotes the operation of permuting indices $1\leftrightarrow2$ and $3 \leftrightarrow 4$, cf.\ equations~\eqref{eq:indexmatching}. Hence there appears to be no real distinction between $\mathfrak{su}_{1/q}(2|2)_\mup{c.e.} \oplus \mathfrak{su}_{q}(2|2)_\mup{c.e.}$ and $\mathfrak{su}_q(2|2)_\mup{c.e.}^{\oplus2}$ symmetry in this case. In our current conventions, up to this basis change on the undotted indices, we have
\begin{equation}
\dsT \simeq \scT(\kappa) \otimes \identity + \identity \otimes \scT(\kappa)
\end{equation}
which is manifestly compatible with $\mathfrak{su}_{q}(2|2)_\mup{c.e.}^{\oplus2}$ symmetry.

Our present results conflict with those of \cite{Arutyunov:2015qva}, whose $\dsT$ matrix only factorizes and satisfies the CYBE after a nonlocal redefinition of the scattering states. Our results agree in the purely bosonic sector, but differ for the fermions.\footnote{To directly compare, note that \cite{Arutyunov:2015qva} defines epsilon with lower indices with the opposite sign from us. Moreover, at the exact S matrix level, the deformation parameter appears to be partially inverted compared to \cite{Arutyunov:2013ega,Arutyunov:2015qva}. We emphasize that this is equivalent to an inconsequential change of basis. In fact, the identification of $q$ depends on the mapping from the exact to the perturbative S-matrix basis in the first place.} As our setups differ throughout the various stages of the computation, it is not straightforward to conclusively determine the origin for this mismatch.\footnote{One possible source lies in the $\kappa$-symmetry gauge fixing, as we were currently unable to compare our gauge choice with the one of \cite{Arutyunov:2015qva}. We thank G.~Arutyunov, R.~Borsato and S.~Frolov for discussions on this point.} Our results show that there exists a gauge choice for which this classically integrable field theory admits a tree-level S matrix that solves the CYBE, which seems like a natural consistency requirement.

\section{Conclusions}

In this paper we studied the worldsheet scattering theory of the light-cone gauge-fixed fermionic $\eta$ deformation of the $\ads$ string. We started by computing the tree-level worldsheet S matrix, showing that it satisfies the classical Yang-Baxter equation and has a factorized structure. Based on expectations regarding the light-cone symmetry algebra, we then determined the exact S matrix factor compatible with $\mathfrak{su}_q(2|2)_\mup{c.e.}$ symmetry for the fermionic Dynkin diagram. By considering the embedding of the light-cone symmetry algebra in the full symmetry algebra relative to the action of the $R$ operator governing the deformation, we found that the two factors of the light-cone symmetry algebra are in fact deformed oppositely, resulting in $\mathfrak{su}_{1/q}(2|2)_\mup{c.e.} \oplus \mathfrak{su}_{q}(2|2)_\mup{c.e.}$ symmetry. The corresponding full exact S matrix is compatible with our tree-level worldsheet computation.

We also revisited the distinguished deformation of $\ads$ in our setup, similarly finding a factorized tree-level $\dsT$ matrix that solves the classical Yang-Baxter equation. In this case the light-cone symmetry algebra is based on the distinguished Dynkin diagram, and it turns out that inversion of the deformation parameter is effectively a symmetry of the exact S matrix factor, so that pragmatically there is no distinction between $\mathfrak{su}_{1/q}(2|2)_\mup{c.e.} \oplus \mathfrak{su}_{q}(2|2)_\mup{c.e.}$ and $\mathfrak{su}_q(2|2)_\mup{c.e.}^{\oplus2}$ symmetry.

There are a number of questions we did not address in this paper. First, it would be interesting to understand what effect the change from distinguished to fermionic deformation has on the exact spectrum of the model, or whether perhaps the models should ultimately be considered equivalent, and if so, how this relates to Weyl invariance. This first of all requires analysis of the corresponding Bethe equations. Second, coming back to the mirror duality mentioned in the introduction, it would be interesting to see whether this feature of the distinguished deformation is also present for our fermionic one. It would not only be interesting to answer this question for the current exact S matrix, but also to investigate the tree-level and exact S matrices for deformations of $\mathrm{AdS}_3 \times \mathrm{S}^3$ where it is possible to realize mirror duality explicitly also in the fermionic sector of the sigma model \cite{Seibold:2019dvf}. Next, for the exact S matrix and Bethe ansatz description of these models it is important to understand the precise identification of the exact parameters $q$ and $\xi$ and the Lagrangian parameters $\kappa$ and $\stringtension$. This is related to questions surrounding quantum corrections to Yang-Baxter deformed backgrounds, where initial studies have thus far focused on $\alpha^\prime$ corrections for homogeneous deformations \cite{Borsato:2019oip,Borsato:2020bqo}, and corrections to (not Weyl invariant) deformed backgrounds to maintain compatibility with RG flow \cite{Hoare:2019ark,Hoare:2019mcc}, see also the very recent \cite{Hassler:2020tvz,Borsato:2020wwk}. At the level of quantum corrections, it would also be great to investigate the one-loop S matrices for both the distinguished and fermionic deformations along the lines of \cite{Roiban:2014cia}, as at loop level we are generically sensitive to Weyl invariance.\footnote{For the distinguished case the one-loop S matrix has been studied using unitarity techniques in \cite{Engelund:2014pla}.} As a lead up to computing such loop corrections, it is moreover important to explicitly check the contributions at quartic order in the fermions at tree level. We could also consider further unimodular inhomogeneous deformations with differing bosonic sectors as in \cite{Delduc:2014kha,Hoare:2016ibq}. In \cite{Hoare:2016ibq} it was shown that the bosonic S matrices of these models are related to the one of the standard inhomogeneous deformation by one-particle momentum-dependent changes of basis. It would be interesting to understand whether this picture continues to hold when including fermions, and if so, to find a matching algebraic picture at the level of the exact S matrix. Finally, it would be very interesting to understand whether the fermionic deformation of the $\ads$ string that we considered here, can be given an interpretation in terms of AdS/CFT.

\section*{Acknowledgments}

We would like to thank Gleb Arutyunov, Riccardo Borsato, Sergey Frolov, Ben Hoare, Alessandro Sfondrini and Linus Wulff for discussions, and Gleb Arutyunov, Riccardo Borsato, Sergey Frolov, Ben Hoare and Arkady Tseytlin for comments on the draft of this paper. The work of FS is supported by the Swiss National Science Foundation through the NCCR SwissMAP. The work of ST and YZ is supported by the German Research Foundation (DFG) via the Emmy Noether program ``Exact Results in Extended Holography''. ST is supported by LT.

\appendix

\section{Spinor conventions}
\label{app:spinorconventions}

Here we briefly set out our conventions regarding the spinors of the Green-Schwarz action and associated objects such as gamma matrices, the vielbein and spin connection. Our conventions are close to those of \cite{Arutyunov:2015qva} but differ in the labeling of coordinates and coset model gamma matrices.

\subsection*{Tangent space}

We introduce tangent space gamma matrices as follows
\begin{equation}
\Gamma_a =
\begin{cases}
\sigma_1 \otimes \gamma_{0} \otimes \mathds{1}_4 & a =0,\\
-\sigma_2 \otimes \mathds{1}_4 \otimes \gamma_{5} & a =1,\\
\sigma_1 \otimes \gamma_{a-1} \otimes \mathds{1}_4 & a =2,3,4,5,\\
-\sigma_2 \otimes \mathds{1}_4 \otimes \gamma_{a-5} & a =6,7,8,9,\\
\end{cases}
\end{equation}
where
\begin{equation}
\begin{aligned}
\gamma_0 &= i \sigma_3 \otimes  \sigma_0 = i \gamma_5,\\
\gamma_1 &= \sigma_2 \otimes \sigma_2,\\
\gamma_2 &= -\sigma_2 \otimes \sigma_1,\\
\gamma_3 &= \sigma_1 \otimes \sigma_0,\\
\gamma_4 &= \sigma_2 \otimes \sigma_3,
\end{aligned}
\end{equation}
are the coset model $\gamma$ matrices of \cite{Arutyunov:2009ga}, and the $\sigma_i$ denote the Pauli matrices, with $\sigma_0 \equiv \mathds{1}_2$. The associated charge conjugation matrix
\begin{equation}
C = i \sigma_2 \otimes K \otimes K, \quad K = -i \sigma_0 \otimes \sigma_2,
\end{equation}
satisfies
\begin{equation}
\Gamma_a^t = - C \Gamma_a C^{-1}, \quad C^t C =\mathds{1}, \quad C^t = - C.
\end{equation}
In these conventions
\begin{equation}
\Gamma_{11} \equiv \Gamma_0 \Gamma_1 \ldots \Gamma_9 = \sigma_3 \otimes \mathds{1}_{16} .
\end{equation}
A Majorana-Weyl spinor satisfies
\begin{equation}
\theta^t C = \bar{\theta} \equiv \theta^\dagger \Gamma^0, \quad \mbox{and} \quad \Gamma_{11} \theta = \theta.
\end{equation}
In the light-cone gauge we fix kappa symmetry as
\begin{equation}
\Gamma^p \theta = 0,
\end{equation}
where we introduce tangent space light-cone coordinates similarly to the curved ones
\begin{equation}
\label{eq:tangentspacelightconeGammas}
\Gamma^p = \tfrac{1}{2}\left(\Gamma^0 + \Gamma^1\right),\quad \Gamma^m = \Gamma^1 - \Gamma^0,
\end{equation}
with labels $p$ and $m$ to distinguish them from the curved space $+$ and $-$.\footnote{In this gauge, any fermion bilinear $\bar{\theta}_i \Gamma^a \ldots \Gamma^e \theta_j$ involving purely transverse tangent space gamma matrices -- those with indices other than $p$ or $m$ ($0$ or $1$) -- is zero.}
We parametrize the components of our two kappa-gauge-fixed Majorana-Weyl spinors as
\begin{equation}
\label{spinor-parametrization}
\theta_1 = \frac{1}{2}
\left(\begin{array}{c}
0\\
0\\
-i \eta_{4\dot2}^\dagger + \eta^{3\dot1}\\
i \eta_{4\dot1}^\dagger + \eta^{3\dot2}\\
0\\
0\\
i \eta_{3\dot2}^\dagger + \eta^{4\dot1}\\
-i \eta_{3\dot1}^\dagger + \eta^{4\dot2}\\
- \theta_{1\dot3}^\dagger - i \theta^{2\dot4}\\
\theta_{2\dot3}^\dagger - i\theta^{1\dot4}\\
0\\
0\\
\theta_{1\dot4}^\dagger - i\theta^{2\dot3}\\
- \theta_{2\dot4}^\dagger - i\theta^{1\dot3}\\
0\\
\vdots\\
0
\end{array}\right), \quad
\theta_2 = \theta_1\bigg\rvert_{\substack{\eta \,\rightarrow \, i \eta\,\,\,\,\\ \theta \,\rightarrow\,-i \theta\\ {}}}
,
\end{equation}
where for the complex-conjugated fields we use the notation $\theta_{a \dot{\alpha}}^\dagger \equiv (\theta^{a \dot{\alpha}})^*$, with $(\theta^{a \dot{\alpha}})^*$ being the complex conjugate of $\theta^{a \dot{\alpha}}$, and similarly for $\eta$.
The index assignment matches the behavior of the components under the $\mathfrak{su}(2)$ transformations of the $Z$ and $Y$ fields of the main text, see equation \eqref{eq:zytoZY}, here in the spinorial representation. This parametrization can be read off by translating the spinor $\theta^{a}_\alpha$ of \cite{Arutyunov:2015qva} to an $8\times8$ matrix using the matrix generators of $\mathfrak{su}(2,2|4)$ used there, and comparing this to the standard parametrization of the fermions in the coset formulation, see e.g.\ equation (1.139) of \cite{Arutyunov:2009ga}.\footnote{In line with appendix C.2 of\cite{Arutyunov:2015qva}, relative to \cite{Arutyunov:2009ga} we permute indices $1$ and $2$, and replace $\theta\rightarrow i \theta$ and $\eta \rightarrow -i\eta$.} In matrix form the kappa gauge $\Gamma^p \theta = 0$ becomes the one used in the coset model formulation, see e.g.\ equation (1.87) of \cite{Arutyunov:2009ga}. Our spinors contain eight complex Grassmann-valued fields: four $\eta$s and four $\theta$s.

\subsection*{Spacetime}

Spacetime $\Gamma$ matrices $\Gamma_M$ are defined as
\begin{equation}
\Gamma_M = e_M^a \Gamma_a,
\end{equation}
where $e$ is the vielbein. In our case the vielbein needs to be chosen appropriately to maintain a straightforward link to coset sigma model objects and associated conventions. Our vielbein is determined by the deformed current $A$ of the sigma model, see eqs.~(2.8) and (2.19) of \cite{Borsato:2016ose}. Taking a coset element appropriate for our $z$ and $y$ variables as in eq.\ (1.147) of \cite{Arutyunov:2009ga}, and evaluating the deformed current, we find
\begin{equation*}
e^{aM} = \left(
\begin{array}{ccccc}
 -\frac{1-\tfrac{z^2}{4}}{1+\tfrac{z^2}{4}} & \kappa  z_1 & \kappa  z_2 & \kappa  z_3 & \kappa  z_4 \\
- \frac{\kappa  z_1}{1+\tfrac{z^2}{4}} & 1-\tfrac{z^2}{4} & -\frac{\kappa  \left(z_3^2+z_4^2\right)}{1-\tfrac{z^2}{4}} & \frac{\kappa  z_2 z_3}{1-\tfrac{z^2}{4}} & \frac{\kappa  z_2 z_4}{1-\tfrac{z^2}{4}} \\
- \frac{\kappa  z_2}{1+\tfrac{z^2}{4}} & \frac{\kappa  \left(z_3^2+z_4^2\right)}{1-\tfrac{z^2}{4}} & 1-\frac{z^2}{4} & -\frac{\kappa  z_1 z_3}{1-\tfrac{z^2}{4}} & -\frac{\kappa  z_1 z_4}{1-\tfrac{z^2}{4}} \\
 -\frac{\kappa  z_3}{1+\tfrac{z^2}{4}} & -\frac{\kappa  z_2 z_3}{1-\tfrac{z^2}{4}} &\frac{\kappa  z_1 z_3}{1-\tfrac{z^2}{4}} & 1-\frac{z^2}{4} & 0 \\
 -\frac{\kappa  z_4}{1+\tfrac{z^2}{4}} & -\frac{\kappa  z_2 z_4}{1-\tfrac{z^2}{4}} & \frac{\kappa  z_1 z_4}{1-\tfrac{z^2}{4}} & 0 & 1-\frac{z^2}{4}
\end{array}
\right)^{aM},
\end{equation*}
for the deformed AdS factor with $a=0,2,3,4,5$ and $M =t,z_1,z_2,z_3,z_4$, and
\begin{equation*}
e^{aM} = \left(
\begin{array}{ccccc}
 \frac{1+\tfrac{y^2}{4}}{1-\tfrac{y^2}{4}} & -\kappa  y_1 & -\kappa  y_2 & -\kappa  y_3 & -\kappa  y_4 \\
 \frac{\kappa  y_1}{1-\tfrac{y^2}{4}} & 1+\tfrac{y^2}{4} & \frac{\kappa  \left(y_3^2+y_4^2\right)}{1+\tfrac{y^2}{4}} & -\frac{\kappa  y_2 y_3}{1+\tfrac{y^2}{4}} & -\frac{\kappa  y_2 y_4}{1+\tfrac{y^2}{4}} \\
 \frac{\kappa  y_2}{1-\tfrac{y^2}{4}} & -\frac{\kappa  \left(y_3^2+y_4^2\right)}{1+\tfrac{y^2}{4}} & 1+\frac{y^2}{4} & \frac{\kappa  y_1 y_3}{1+\tfrac{y^2}{4}} & \frac{\kappa  y_1 y_4}{1+\tfrac{y^2}{4}} \\
 \frac{\kappa  y_3}{1-\tfrac{y^2}{4}} & \frac{\kappa  y_2 y_3}{1+\tfrac{y^2}{4}} & -\frac{\kappa  y_1 y_3}{1+\tfrac{y^2}{4}} & 1+\frac{y^2}{4} & 0 \\
 \frac{\kappa  y_4}{1-\tfrac{y^2}{4}} & \frac{\kappa  y_2 y_4}{1+\tfrac{y^2}{4}} & -\frac{\kappa  y_1 y_4}{1+\tfrac{y^2}{4}} & 0 & 1+\frac{y^2}{4}
\end{array}
\right)^{aM},
\end{equation*}
for the deformed sphere factor with $a=1,6,7,8,9$ and $M =\phi,y_1,y_2,y_3,y_4$. Other components of the vielbein vanish. We raised the curved index to get more compact expressions. At $\kappa=0$ the vielbein is diagonal and associates tangent indices to coordinates as
\begin{equation}
\begin{pmatrix}
0 & 1 & 2 & 3 & 4 & 5 & 6 & 7 & 8 & 9\\
t & \phi & z_1 & z_2 & z_3 & z_4 & y_1 & y_2 & y_3 & y_4
\end{pmatrix}
\end{equation}
The spin connection can be similarly extracted, however it is not independent and can also be found via
\begin{equation}
\omega^{ab}_M = -2e^{[a|N} \partial_{[M} e^{|b]}_{N]} - e^{a P} e^{b Q} \partial_{[Q} e^c_{P]} e_{cM}.
\end{equation}

\section{Feynman diagrammatics}
\label{feynman-diagrammatics}
We used standard Feynman diagram methods to determine the perturbative T matrix --
with the two major steps being the calculation of the Feynman rules and Feynman amplitudes.
We performed these two procedures in Mathematica, using the packages \emph{FeynRules} \cite{Alloul:2013bka} and \emph{FeynArts} \cite{Hahn:2000kx} respectively.

This section states the implementation details and highlights certain issues arising from the intricacies of the model at hand.
In particular it has scalar valued fermions, which further are complex off shell, but become real on shell (see the end of \cref{sec-mode-expansion}).
The bosons in turn are always constrained to be real by the reality condition \cref{reality-bosons}.

\paragraph{Feynman rules}
We describe our model as 8 real bosons and 8 complex fermions
and will take the (on-shell) reality conditions into account only when calculating the amplitudes in the second step.

When turning the interaction terms into vertex factors the FeynRules package assumes a $(+,-)$ signature and therefore prefactors each term with $+\I$.
Additionally, it replaces derivatives by components of the covariant momentum: $(\partial_\tau, \partial_\sigma) \to -\I (\omega, p)$.
We want to adopt a $(-,+)$ signature convention however, which requires a prefactor of $-\I$,
and moreover, we want to use the components of the contravariant momentum vector, i.e.~replace $(\partial_\tau, \partial_\sigma) \to \I (-\omega, p)$.
To match these two choices we add an extra sign to each vertex and replace $p \to -p$ in the output of FeynRules.

For the scalar fermions we follow the algorithm for Feynman diagrams with general fermionic fields \cite{Denner:1992vza}.
In particular, this requires us to keep track of a fermion flow direction for each vertex involving fermions.
It is not sufficient to simply look at the particle/anti-particle flow (like can be done for Dirac fermions), because certain interaction terms with fermionic parts break this flow. (For example terms proportional to $\theta \theta$ or $\theta^\dagger \theta^\dagger$.)
FeynRules is interfering with the proper tracking of the fermion flow by not respecting the ordering of fermionic fields in the input and bringing them into alphabetic order internally.
We were able to resolve this issue by ordering the input already before giving it to the package, adding extra signs when anticommuting fermions.

Our light cone gauge explicitly breaks the Lorentz invariance of the interaction terms.
To support this, we had to perform minor modifications to the code of FeynRules.
Further, we uncovered two bugs in version 2.3.36 of the package, which caused certain vertices with momentum dependence to be dropped from the output.
We reported these to the developers and proposed a fix.
A patch file\footnote{Apply with \texttt{patch -p0 < FeynRules.patch}.} containing the modifications and the bug fixes is attached to the arXiv submission of this paper.

Finally, we obtain all the 4-point vertices coming from the interaction Lagrangian and can use them to calculate the Feynman amplitudes.

\paragraph{Feynman amplitudes}
With the 4-point vertices at hand we can determine the amplitudes for the $2 \to 2$ scattering described in \cref{T-matrix}.
The FeynArts package also assumes a $(-,+)$ signature and therefore prefactors each amplitude with $-\I$.
Our choice of $(+,-)$ requires a prefactor of $+\I$ however.
We add an extra sign to each amplitude thus.
The in- and out-states are assigned according to the occurrence of the operators $a^{M \dot{N}}$ and $a^\dagger_{M \dot{N}}$ in the mode expansion in eqs.~\eqref{mode-expansion}.
To account for the on-shell reality of the fermions, we sum the contributions from their fields and anti-fields.
Further data taken from the mode expansion is the dispersion relation for $\omega_p$, the prefactors $\frac{1}{2 \sqrt{\omega_p}}$, $-\frac{1}{2 \sqrt{\omega_p}}$ and $\frac{1}{\sqrt{\omega_p}}$ for incoming bosons, outgoing bosons and fermions respectively, and the fermionic wave functions (in the notation of figure~2.3 of \cite{Denner:1992vza})
\begin{align*}
	u_\theta(p) & = -\I \E^{-\I\pi/4} f_p^*
	&
	u_\eta(p) & = +\I \E^{-\I\pi/4} f_p
	\\
	\bar{v}_\theta(p) & = +\I \E^{+\I\pi/4} h_p
	&
	\bar{v}_\eta(p) & = -\I \E^{+\I\pi/4} h_p^*
	\\
	\bar{u}_\theta(p) & = -\I \E^{-\I\pi/4} h_p^*
	&
	\bar{u}_\eta(p) & = +\I \E^{-\I\pi/4} h_p
	\\
	v_\theta(p) & = +\I \E^{+\I\pi/4} f_p
	&
	v_\eta(p) & = -\I \E^{+\I\pi/4} f_p^*
\end{align*}

FeynArts already implements the algorithm described in \cite{Denner:1992vza} for Feynman diagrams containing general fermionic fields and only requires the reversed-fermion-flow vertices $\Gamma'$ and wave functions $u'$,~$v'$ as input.
In contrast to the case discussed in \cite{Denner:1992vza}, the fermions at hand are scalars and the reversed quantities are therefore given by anticommuting Grassmann fields.
This simply adds extra minus signs,
\begin{equation}
	\Gamma' = -\Gamma
	\mcomma \quad
	u' = -u
	\mcomma \quad
	v' = -v
	\mcomma
\end{equation}
and similarly for the barred versions.
$\Gamma$ represents the forward-fermion-flow vertices, see figure~2.1 of \cite{Denner:1992vza}.

A further modification stems from the fact that FeynArts orders the particles in the in-state opposite to how the existing literature does and how we present them.
To account for this we have to add an extra minus sign for amplitudes with two fermionic in-states.

\newcommand{\FeynAmplitude}{\mathcal{M}}
Finally, our sought for $\dsT$ is related to the modified FeynArts amplitudes $\FeynAmplitude$ by
\begin{equation}
	\begin{aligned}
		\dsT(p_1, p_2) & =
			\int \dd{k_1} \dd{k_2} \delta(p_1 + p_2 - k_1 - k_2) \delta(\omega_{p_1} + \omega_{p_2} - \omega_{k_1} - \omega_{k_2}) \; \FeynAmplitude(p_1, p_2, k_1, k_2)
		\\ & =
			\frac{\omega_{p_1} \omega_{p_2}}{\abs{p_1 \omega_{p_2} - p_2 \omega_{p_1}}} (\FeynAmplitude(p_1, p_2, p_1, p_2) + \FeynAmplitude(p_1, p_2, p_2, p_1))
		\mdot
	\end{aligned}
\end{equation}
Here $k_1$ and $k_2$ denote the momenta of the outgoing particles.
Due to energy and momentum conservation these are restricted to take on the same values as the incoming momenta $p_1$ and $p_2$.
To follow the existing literature we assume that $p_1 > p_2$.

\section{\texorpdfstring{$\mathfrak{su}(2|2)$ $R$ operators}{𝔰𝔲(2|2) R operators}}
\label{app:su22Roperators}

Here we derive the precise form of the $R$ operators corresponding to the fermionic and distinguished $q$ deformations of $\mathfrak{su}(2|2)$ used in section \ref{sec:exactSmatrix} and \cite{Beisert:2008tw}, and express them in a basis of the form
\begin{equation}
\label{eq:su22form}
\left(
\begin{array}{cc}
\mathbb{R} & -\mathbb{Q}^\dagger \\
\mathbb{Q} & \mathbb{L}
\end{array}
\right),
\end{equation}
referred to in sections \ref{sec:comparison} and \ref{sec:distinguishedcase}. As the fermionic case is built on the distinguished case, we first consider the latter.

\paragraph{Distinguished deformation.} We start with \crefrange{eq:concreterootaction}{eq:rootmatrixrealization}, taking $\aalt=\balt=\calt=\dalt=1$ for unitarity in the undeformed limit, and $a=d=1$ and $b=c=0$ for the standard fundamental representation of $\mathfrak{su}(2|2)$ with $C=+1/2$. Note the anti-canonical ordering of $|\psi_1\rangle$ and $|\psi_2\rangle$ with respect to the basis vectors in eqs.~\eqref{eq:su22basisvectors}. We have
\begin{equation}
\mathbb{E}_1 =
\begin{pmatrix}
0& 0 & 0 & 0 \\
1& 0 & 0 & 0 \\
0& 0 & 0 & 0 \\
0& 0 & 0 & 0 \\
\end{pmatrix}, \quad
\mathbb{E}_2 =
\begin{pmatrix}
0& 0 & 0 & 0 \\
0& 0 & 0 & 0 \\
0& 1 & 0 & 0 \\
0& 0 & 0 & 0 \\
\end{pmatrix},\quad
\mathbb{E}_3 =
\begin{pmatrix}
0& 0 & 0 & 0 \\
0& 0 & 0 & 0 \\
0& 0 & 0 & 0 \\
0& 0 & 1 & 0 \\
\end{pmatrix},
\end{equation}
while the remaining non-simple positive roots can be obtained by repeated commutators of these. In this case all positive roots are lower diagonal, and all negative roots are upper diagonal. The $R$ operator that acts as multiplication by $-i$ on the positive roots, $+i$ on the negative roots, and $0$ on the Cartan generators, is then given by
\begin{equation}
R_{\mathfrak{su}(2|2)}(M)_{ij} = - i \epsilon_{ij} M_{ij} \mcomma \qquad \epsilon =
\left(
\begin{array}{cccc}
0 & -1 & -1 & -1 \\
+1 & 0 & -1 & -1\\
+1 & +1 & 0 & -1\\
+1 & +1 & +1 & 0
\end{array}
\right).
\end{equation}
This is the form the $R$ operator takes on an algebra element of the form \eqref{eq:su22form}. To see this, we note that in \cite{Beisert:2008tw} the simple positive roots are taken to be $\mathfrak{R}^{2}{}_1$, $\mathfrak{Q}^{2}{}_2$ and $\mathfrak{L}^{1}{}_2$. Taking into account \eqref{eq:indexmatching}, these generators can be related to ours as $\mathfrak{R}^{\alpha}{}_\beta \sim \mathbb{L}_{\alpha}{}^\beta$, $\mathfrak{L}^{a}{}_b \sim \mathbb{R}^a{}_b$, and $\mathfrak{Q}^\alpha{}_b = \mathbb{Q}_\alpha{}^b$, in line with the grading of their indices, and the algebra relations. With these identifications,\footnote{Recall the relative index permutation of indices $1$ and $2$ (see the end of \cref{sec:expansion-of-action}) when comparing to e.g.\ \cite{Arutyunov:2009ga}. This index permutation on the Latin (bosonic) indices can be viewed as a counterpart to the different index positions on the generators of \cite{Beisert:2008tw} and ours, since raising and lowering indices with the two dimensional $\epsilon$ tensors effectively permutes the indices. Relatedly, we did not do a permutation of the Greek (fermionic) indices with respect to \cite{Arutyunov:2009ga}, matching the anti-canonical ordering of the $\psi$ indices of \cite{Beisert:2008tw} with respect to the matrix basis, see \cref{eq:su22basisvectors}.}  the positive simple roots given above match with the matrix structure \eqref{eq:su22form} of the algebra element.

\paragraph{Fermionic deformation.} We can use the Lusztig transformation of eqs.~\eqref{eq:LusztigTransform} to determine our new simple roots, and commutators for the remainder. Demanding the usual action of $R$ on these roots then gives
\begin{equation}
R_{\mathfrak{su}(2|2)}(M)_{ij} = - i \epsilon_{ij} M_{ij} \mcomma \qquad \epsilon =
\left(
\begin{array}{cccc}
0 & -1 & -1 & -1 \\
+1 & 0 & +1 & -1\\
+1 & -1 & 0 & -1\\
+1 & +1 & +1 & 0
\end{array}
\right).
\end{equation}

\pdfbookmark[section]{\refname}{bib} 
\begin{sloppypar} 
\bibliographystyle{nb}
\bibliography{bibliography}
\end{sloppypar}

\end{document}